\documentclass[final,5p]{elsarticle}
\usepackage[utf8x]{inputenc}
\usepackage{hyperref}
\usepackage{amssymb,amsmath}

\journal{Astronomy and Computing}

\biboptions{authoryear}

\begin{document}

\begin{frontmatter}

\title{Modeling Circumstellar Disc Fragmentation and Episodic Protostellar Accretion with Smoothed Particle Hydrodynamics in Cell}

\author[nsu,mymainaddress]{Olga P. Stoyanovskaya\corref{mycorrespondingauthor}}
\ead{stop@catalysis.ru}

\author[mysecondaryaddress]{Nikolay V. Snytnikov}

\author[nsu,mymainaddress]{Valeriy N. Snytnikov}
\cortext[mycorrespondingauthor]{Corresponding author}

\address[nsu]{Novosibirsk State University, Pirogova, 2 Novosibirsk, 630090, Russia}
\address[mymainaddress]{Boreskov Institute of Catalysis, Pr.Academika Lavrentieva, 5, 630090, Novosibirsk, Russia, }
\address[mysecondaryaddress]{Institute of Computational Mathematics and Mathematical Geophysics, Lavrentieva, 6, Novosibirsk, 630090, Russia}

\begin{abstract}

We discuss the ability of the smoothed particle hydrodynamics (SPH) method combined with a grid-based solver for the Poisson equation to model mass accretion onto protostars in gravitationally unstable protostellar discs. We scrutinize important features of coupling the SPH with grid-based solvers and numerical issues associated with (1) large number of SPH neighbours and (2) relation between gravitational softening and hydrodynamic smoothing length. 

We report results of our simulations of razor-thin disc prone to fragmentation and demonstrate that the algorithm being simple and homogeneous captures the target physical processes - disc gravitational fragmentation and accretion of gas onto the protostar caused by inward migration of dense clumps. 

In particular, we obtain two types of accretion bursts: a short-duration one caused by a quick inward migration of the clump, previously reported in the literature, and the prolonged one caused by the clump lingering at radial distances on the order of 15-25~au. The latter is culminated with a sharp accretion surge caused by the clump ultimately falling on the protostar. 

\end{abstract}


\end{frontmatter}

\section{Introduction} \label{sec:intro}

Dynamical processes in accretion discs of young stellar objects has recently gained much attention. Among them are short-lived episodes of high-rate accretion of FU-Orionis-type stars (FUors). While there exist many theoretical models that can explain FUors (see e.g. a review by \citet{Audard}), the disc fragmentation model proposed by \citet{VorobyovBasu2006, VorobyovBasu2010, VorobyovBasu2015} has a certain appeal because it suggests a causal link between episodic accretion in young protostars and the initial stages of planet formation in gravitationally unstable discs. In this model, accretion and luminosity bursts are the result of massive fragments forming in the outer parts of gravitationally unstable discs and spiraling onto the star owing to the loss of angular momentum via gravitational interaction with spiral arms and other fragments. If fragments have enough time to accumulate solid protoplanetary cores deep in their interiors, this mechanism can propose an alternative gateway to the formation of planets via the so-called tidal downsizing hypothesis \citep{Nayakshin, Boley}.

Mathematical models of gravitational fragmentation of razor-thin and thick discs were applied to investigate episodic protostellar accretion e.g. by  \cite{VorobyovBasu2006,MachidaAc}. Numerical simulations of protostellar discs prone to fragmentation require a high numerical resolution to resolve the minimum perturbation unstable to growth under self-gravity (e.g. \citet{Truelove}). 

That is why Lagrangian methods, such as the smoothed particle hydrodynamics \citep{Lucy,GMSPH} where resolution naturally follows mass are often adopted for such simulations. 

For SPH simulations of self-gravitating gaseous disc calculation of short-term and long-term forces is optimized to avoid looping over all particle pairs. In particular, to compute short-range hydrodynamic forces near-neighbour particles are determined. To compute long-range gravitational force the gravity of a distant group of particles is substituted by the gravity of one particle of the total mass. It means that for efficient forces calculation in SPH we organize kind of Euler decomposition for particles. Moreover, for simulation of self-gravitating gas with SPH it is natural to perform the decomposition once, and than to adopt it twice for short-range and long-range force calculation. During last several decades an approach when tree-code is used to determine nearest neigbours and to calculate disc self-gravity in SPH proved to be the optimal choice for serial and high-performance computing e.g. \citet{SpringelSPH}.

On the other hand, due to fast development of supercomputer architecture workstations with different power and number of processors become available for numerical simulations. This diversity of available supercomputers promotes interest to the algorithms that could be easily transferred from small clusters with several computational nodes to large-scale machines with thousand nodes. This work was motivated by the idea that logically simple algorithms can be transferred to large-scale supercomputers more efficiently than complex branching algorithms, albeit they could demonstrate inferior performance on serial and medium-sized supercomputers. For this reason we searched a logical simplification of a standard approach to organize SPH simulations of self-gravitating gaseous discs. 

It is well known, that after substitution of arbitrarily spaced masses with equigravitating set of masses located on uniform Cartesian grid the gravitational potential could be found faster due to convolution theorem. It allows us to propose an algorithm that combines the SPH technique and a grid-based method for calculating gravitational forces to solve numerically the Euler equations for the gas dynamics. This modification to the standard SPH approach, wherein the gravity force is calculated using a tree-code, increases simplicity and homogeneity of the numerical scheme, but features one of two inevitable negative aspects: inequality between gravitational softening and hydrodynamic smoothing length \citep{Nelson} or --- in case when the smoothing length is kept fixed and equal to gravitational softening length --- usage of smaller or larger than optimal number of neighbors in SPH \citep{Price,Wendland}. 

As a first step we investigate the scheme when the gravitational softening and hydrodynamics smoothing length are kept equal. A feature of usage SPH instead of grid-based gas dynamics on uniform mesh is higher actual resolution of hydrodynamic parameters of formed clumps that are small-scaled anticyclone vortices with high velocity, density and pressure gradient. We aim to demonstrate that our scheme (1) provides results that are independent on numerical resolution when the dynamics of fragmenting disc is simulated, (2) allows to capture different regimes of accretion of gas onto the protostar in dense discs. We report our preliminary results in detail focusing on (a) confirmation of already described scenarios for accretion bursts using numerical methods that differ from those used in the previous studies by \citet{VorobyovBasu2015,MachidaAc} and (b) finding new modes of episodic accretion.   

The paper is organized as follows. In Section~\ref{sec:equations} we presented the model of self-gravitating accretion disc used for the simulations. In Section~\ref{sec:methods} we described the numerical model focusing on the way of coupling SPH and mesh. In Section~\ref{sec:tests} we focused on problematic numerical issues of protostellar accretion simulation using SPH and presented results of simulations, where we test specially designed by \citet{TC1992,Wendland} measures to suppress clumping instability for the case when more than optimal number of neighbors in SPH should be used. In Section~\ref{sec:ResGeneral} we compared numerical results obtained for our disc model with different numerical resolution, especially focusing on the ratio between hydrodynamic smoothing length and gravitational softening length, which is known as a possible source of numerical artifacts in the solution e.g. \citet{Nelson}. In Section~\ref{sec:results} we presented results of modeling accretion for the fragmenting and non-fragmenting discs. 

\section{Basic equations}
\label{sec:equations}
The computational experiments reported in this paper were carried out within a razor-thin model of the disc. This means that we neglected the vertical motion of matter and considered the dynamics of the disc where its entire mass was concentrated inside the equatorial plane of the system. 

Since we do not focus on the thermal dynamics of fragments, we treat cooling via simple assumption about the equation of state. We used adiabatic evolution where specific entropy is held fixed and entropy generation in shocks is ignored (also used e.g. by \citet{Pickett1998,Pickett2000}). More details on classification of cooling models of the discs can be found in \citep{Durisen}. We note, that this approach allows to mimic the temperature of migrating clumps (see Appendix) obtained from simulations of other authors \citep{ZhuClumps,NayakshinCha,Vorobyov2013}.

The gas component was described by the following gas dynamics
equations:
\begin{equation}
\label{EUcont}
\displaystyle\frac{\partial \Sigma}{\partial t}+div(\Sigma
\textbf{\textit{v}})=0, \\
\end{equation}

\begin{equation}
\label{EUmotion}
\displaystyle\Sigma \frac{\partial \textbf{\textit{v}}}{\partial
t}+\Sigma(\textbf{\textit{v}} \cdot \nabla)\textbf{\textit{v}}=-\nabla
p^* - \Sigma\nabla\Phi_{sum}, \\
\end{equation}

\begin{equation}
\label{EUent}
\displaystyle\frac{\partial S^*}{\partial t}+(\textbf{\textit{v}} \cdot \nabla) S^* =
0, \ \ \ \ p^*=T^*\Sigma. \ \ \
\end{equation}

These gas dynamics equations include surface quantities that were
obtained from volume quantities by integration with respect to the
vertical coordinate $z$:
\[\Sigma=\int_{-\infty}^{+\infty} \rho dz; \ \ \
p^*=\int_{-\infty}^{+\infty} p dz.\]

Here, $\textbf{\textit{v}}=(v_x,v_y)$ is the two-component gas
velocity, and  $p^*$ is the surface gas pressure.
$T^*=\displaystyle\frac{p^*}{\Sigma}$, $S^* = \ln
\displaystyle\frac{T^*}{\Sigma^{\gamma^*-1}}$ are the quantities
similar to gas temperature and entropy. $\gamma^*$ is a 2D version
of $\gamma$ \citep{Fridman}, which is related to the
constant ratio of specific heats as $\gamma^*=3-\displaystyle\frac{2}{\gamma}$. 

$\Phi_{sum}$ is the gravitational potential in which the motion occurs, defined as the sum of central body potential and disc potential:
$$
\Phi_{sum}=\Phi_c+\Phi, \Phi_c=-\displaystyle\frac{M_c G}{r},
$$
where $M_c$  is the mass of central body. $\Phi$ is the potential of self-consistent gravitational field, which satisfies Poisson equation:
$$
\Delta\Phi=4 \pi G \Sigma, \ \ \ \Phi\longrightarrow_{r\rightarrow \infty} 0.
$$

\subsection{Notions of gravitational instability theory used in the paper}

The dispersion relation for the considered model of razor-thin disc is introduced in several works (see e.g. \citet{Book,Nelson})
$$\omega ^2=c_s^2 k^2+ \kappa ^2 -2 \pi G \Sigma |k|, $$ where
$\kappa $ is the epicyclic frequency, and $c_s$ is the sound speed. For keplerian discs $ \kappa =\Omega $. 

For an extended hypothetical sheet of gas $\kappa =0$ and $\omega ^2=c_s^2 k^2 - 2 \pi G \Sigma |k|,$ from which one can obtain the critical Jeans length $\lambda_J$:
$$k_J=\displaystyle\frac{2 \pi}{\lambda_J}=\frac{2 \pi G \Sigma}{c_s^2}, \ \ \lambda_J=\displaystyle\frac{c_s^2}{G \Sigma}.$$

For the rotating disc, one can obtain the Toomre condition of marginal stability from the equations $\displaystyle \frac{d \omega^2}{dk}=0, \ \ \omega^2=0:$
$$k_T=\frac{\pi G \Sigma}{c_s^2}; \ \ \lambda_T=\frac{2 \pi}{k_T}=2 \lambda_J.$$
By substitution of the found value $k_T$ into $\omega^2=0$ we derive the critical value of Toomre parameter $Q=\displaystyle\frac{\Omega c_s}{\pi G \Sigma}=1$. For $Q>1$ the razor-thin disc is stable against growth of radial perturbations, for $Q<1$ the disc is unstable against growth of radial perturbations.

\subsection{Initial conditions}
\label{sec:init}

The simulated disc had inner radius $R_{min}=10$~au and outer radius $R_{max}=100$~au. The surface temperature and density of the disc were specified at zero time. Based on the results of simulation by \citet{Vorobyov2010}, in the calculations presented in this paper the initial density of the gas was taken as $\displaystyle\Sigma=\Sigma_0\frac{1}{r}$, where $\Sigma_0$ is found from the equation $2 \pi \displaystyle\int_{R_{min}}^{R_{max}} \Sigma_0 dr =M_{disc}$. The gas temperature at zero time was specified as  $T=\displaystyle\frac{T_0}{\sqrt{r}}$, where $T_0$ is a user-defined parameter. We used $\gamma^*=1.4$.

The gas velocity was determined from an equilibrium between centrifugal and centripetal gravitational forces:
$\displaystyle\frac{v_{\phi}^2}{r}=\frac{1}{\Sigma}\frac{\partial
p^*}{\partial r}+\frac{\partial \Phi_{sum}}{\partial r},$
$v_r=0$.

\begin{figure}
  \includegraphics[width=\columnwidth]{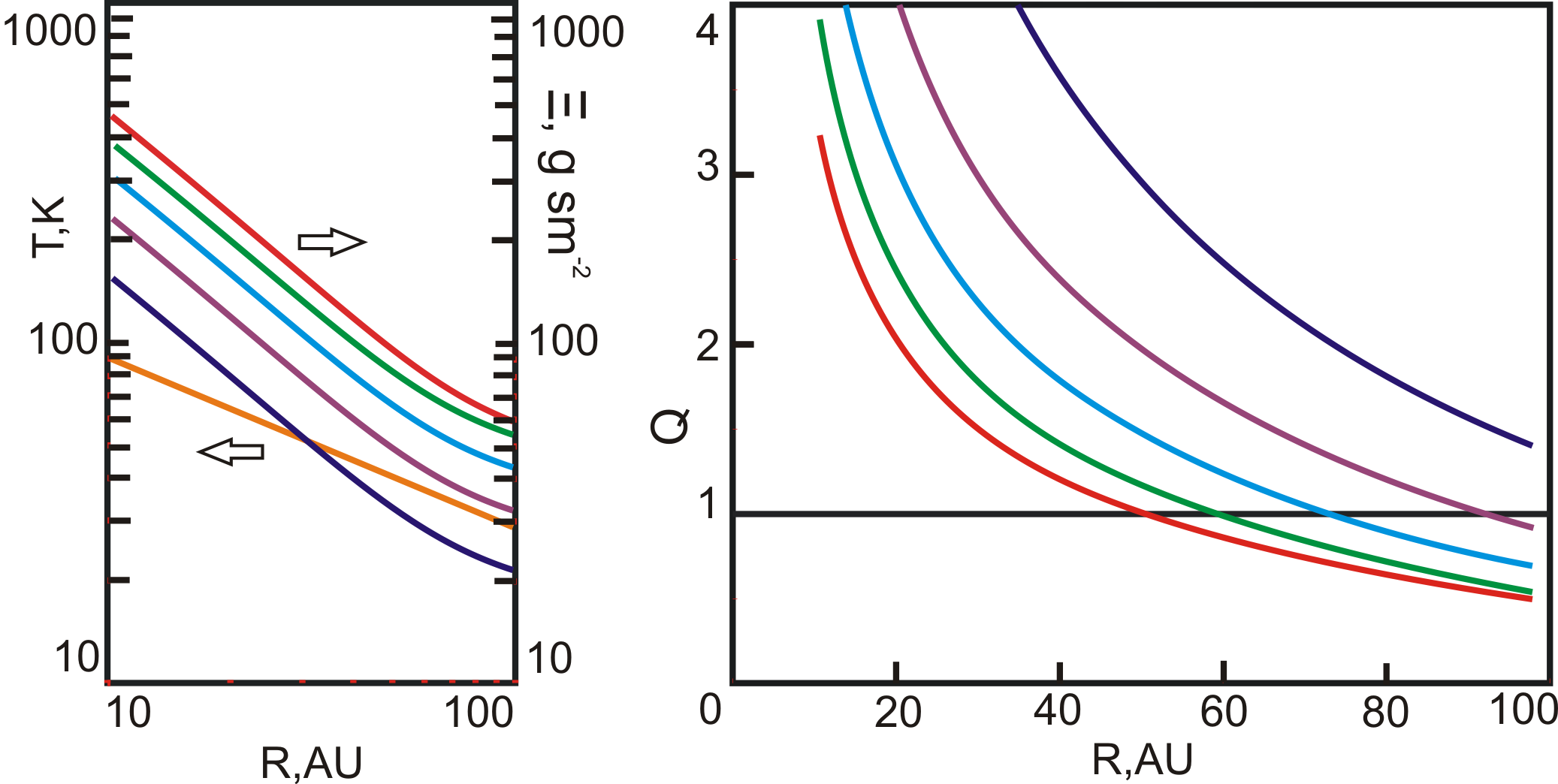} 
  \caption{Initial temperature (for all models orange, left panel), surface density (left panel) and Toomre parameter Q distribution (right panel) for the discs of 0.1 (Model 1, deep blue), 0.15 (Model 2, pink), 0.2 (Model 3, blue), 0.25 (Model 4, green), and 0.3 (Model 5, red) Solar masses. }
 \label{fig:init}
 \end{figure}

For our simulation we used five initial disc setups that differ only in the mass of the disc. Initial temperature was taken equal to 100 K at 10 au and about 30 K at 100 au. The mass of the protostar at zero time is equal to 0.8 Solar mass. Initial disc mass was taken from 0.1 Solar mass (model 1) to 0.3 Solar mass (model 5). Initial temperature and surface density distribution, and the obtained value of initial Toomre parameter are given on Fig.\ref{fig:init}. Each setup was calculated several times: differences were in the number of SPH particles and in the form of kernel. The extensive list of runs is given in Table 1.

\section{Numerical methods}
\label{sec:methods}

We carried out a computer simulation of protoplanetary disc dynamics using a code based on the method of splitting with respect to the involved physical processes. The gas dynamics equations and Poisson equation were solved at each time step. 

\subsection{SPH setup}
\label{sec:SPH}

The gas dynamics equations were solved using the SPH method \citep{SPH}. The SPH calculation formulas implemented in the code were obtained from equations \ref{EUcont}  --- \ref{EUent} written in the Lagrangian form:
\[ \frac{d
\Sigma}{dt}=\Sigma \cdot {div}{\textbf{v}}, \ \
\frac{d\textbf{v}}{dt}=-\frac{1}{\Sigma} \nabla p - \nabla \Phi_{sum}, \
\ \frac{d\textbf{r}}{dt}=\textbf{v}, \ \ \frac{dS}{dt}=0,
\] where $\displaystyle\frac{d}{dt}=\frac{\partial}{\partial t}+ v \cdot
\nabla$.

In our calculations three different kernels were used: (1) the cubic spline for 2D space $W$:
\begin{equation}
\label{eq:CubicSpline}
W(q,h)=\frac{5}{14 \pi h^2}\left\{ \begin{array}{l}
                 [(2-q)^3-4(1-q)^3],  \ \ 0\leq q \leq 1,\\
                 \ \ \ \ \ \ [2-q]^3,  \ \ \ \ \ \ \ \ \ \ \ \ \ \ 1 \leq q \leq 2, \\
                 0, \ \ \ \ \ \ \ \ \ \ \ \ \ \ \ \ \ \ \ \ \ \ \ \ \ \ \ \ \ \  \ \ \ q > 2,
                 \end{array} \right.
\end{equation}
(2) its \citet{TC1992} modification (TC) in a form 
\begin{equation}
\label{eq:ThomasCouchman}
W'_*(q,h)=\frac{5}{14 \pi h^2}\left\{ \begin{array}{l}
                 -4,  \ \ \ \ \ \ \ \ 0\leq q \leq q_0,\\
                 -3(2-q)^2+12(1-q)^2,  \ \ \ \ q_0 \leq q \leq 1,\\
                 \ \ \ \ \ \ -3(2-q)^2,  \ \ \ \ \ \ \ \ \ \ \ \ \ \ 1 \leq q \leq 2, \\
                 \ \ \ \ \ \ \ 0, \ \ \ \ \ \ \ \ \ \ \ \ \ \ \  \ \ \ \ \  \ \ \ q > 2,
                 \end{array} \right.
\end{equation} and (3) quintic Wendland function taken from the paper by \citet{Wendland} in a form
\begin{equation}
\label{eq:WendlandKernel}
W(q,h)=\frac{7}{64 \pi h^2} \left\{ \begin{array}{l}
                  (2-q)^4(1+2q),   \ \ \ \ \ \ \ 0 \leq q \leq 2, \\
                  \ \ \ 0, \ \ \ \ \ \ \ \ \ \ \ \ \ \ \  \ \ \ \ \  \ \ \ q > 2.
                 \end{array} \right.
\end{equation}
where $q=\displaystyle\frac{|\textbf{x}|}{h},$ $\textbf{x}$ is the 
radius vector of a space point, $\displaystyle q_0=\frac{2}{3}$.

The surface density of the gas where the \textit{i}th particle resides was calculated as the sum $\Sigma_i = m \sum_{j=1}^{N} W_{ij},$ where $N$ was the number of simulated SPH particles, m - the mass of the SPH-particle. 

In our calculations we used constant smoothing length equal to the linear size of cartesian grid cell $h_{SPH}=h=h_{grid}$. 

The equation of motion \ref{EUmotion} was approximated so that the impulse and angular momentum were preserved and artificial viscid force was added:
\[
\frac{d \textbf{v}_i}{dt}=- \sum_j
m_j(\frac{p^*_j}{\Sigma_j^2}+\frac{p^*_i}{\Sigma_i^2}+\Pi_{ij})
\nabla_i W_{ij} -\textbf{F}_i, \]

\[ \Pi_{ij}=\left\{ \begin{array}{l}
                 \displaystyle\frac{-\alpha \overline{c_{ij}} \mu_{ij}+\beta \mu_{ij}^2}{\overline{\Sigma_{ij}}},  \ \ v_{ij}r_{ij}<0,\\
                 0, \ \ \ \ \ \ \ \ \ \ \ \ \ \ \ \ \ \ \ \ \  \ v_{ij}r_{ij}\geq 0,,
                 \end{array} \right. \]
\[
\mu_{ij}=\frac{h v_{ij}r_{ij}}{|r_{ij}|^2+0.01h^2},
 \ \overline{c_{ij}}=\frac{1}{2}(c_i+c_j), \ \overline{\Sigma_{ij}}=\frac{1}{2}(\Sigma_i+\Sigma_j),
\]
\[
 v_{ij}=v_i-v_j, r_{ij}=r_i-r_j,
\displaystyle\textbf{F}_i=\nabla \Phi_i + \frac{M_c G}{r_i^3}\textbf{x}.
\] 

For calculation formulas we used the notations:
\[
W_{ij}=W(|r_i-r_j|,h), \ \ \nabla_i
W_{ij}=\displaystyle\frac{x_i-x_j}{h}\frac{\partial
W_{ij}}{\partial q}.
\]

We used a standard artificial viscosity with parameters $\alpha=1, \ \beta=1$ \citep{SPH}. In our model we have to reproduce accurately a supersonic ($Ma>14$, where $Ma$ is Mach number) shear flow of the inner part of the disc, which is non-trivial for SPH due to the following reason. Adding artificial viscosity means adding the pressure-correction term into the motion equation; to keep the balance of energy we then have to add the term responsible for kinetic to inner energy transfer into the equation of energy. For the case of supersonic shear flow, this term (artificial heating) provides a significant increase in gas temperature and a transition of its flow from supersonic to subsonic in the inner part of the disc. E.g. for our test calculation of the kinetic energy, a loss due to viscosity on $128 \times 128$ grid cells and 40000 SPH particles is about 10 per cent, when transferring this energy into internal means more than a twofold overestimation of temperature. The simplest way to treat this problem is to add artificial viscosity into the energy equation only and allow the system to undergo a slow energy loss (cooling). 

For nearest neighbour search we used the linked-list algorithm where at every time step the particles were assorted into uniform Cartesian grid cells. The most suitable for such assortment is the counting sort algorithm that has a linear complexity.  

\subsection{Gravitational potential calculation}
\label{sec:potential}
To compute three-dimensional gravitational potential we used the convolution method \citep{Hockney, eastwood}. 
Instead of solving the Dirichlet problem for Poisson equation (\ref{eq:poisson}) with a boundary condition defined for the 3D infinite domain:
\begin{equation}
\label{eq:poisson}
\displaystyle
\begin{array}{l}
\Delta \Phi(\mathbf{x}) = 4 \pi G \rho(\mathbf{x})=4 \pi G \Sigma, \\
\Phi(\mathbf{x})|_{\mathbf{x} \to \infty} = 0,
\end{array}
\end{equation}
the method makes use of fast calculation of the fundamental solution for Poisson equation:
\begin{equation}
\label{eq:poisson_integral}
\displaystyle
\Phi(\mathbf{x_0}) = -\int\frac{G \rho(\mathbf{x})d\mathbf{x}}{|\mathbf{x_0}-\mathbf{x}|}.
\end{equation}
For the Cartesian uniform grid with the number of nodes  $N_x \times N_y \times N_z$ 
and spatial grid steps $h_x, h_y, h_z$, the equation reads:
\begin{eqnarray}
\label{eq:poisson_integral_discrete}
\displaystyle
\Phi(x_0, y_0, z_0) = \\
\displaystyle
-\sum\limits_{i=1}^{N_x-1}\sum\limits_{j=1}^{N_y-1}\sum\limits_{k=1}^{N_z-1}
\frac{q_{i,j,k}}{\sqrt{(x_i-x_0)^2+(y_i-y_0)^2+(z_i-z_0)^2}},
\end{eqnarray}
where $q_{i,j,k} = G \rho_{i,j,k} \cdot (h_x h_y h_z)$~ is a point mass located in the grid node $(i,j,k)$.

To compute forces we need only those values of 3D gravitational potential that are located in the plane $z=z_0$. 
And there is no need to compute and store the other grid values. It reduces~(\ref{eq:poisson_integral_discrete})
to the following expression:
\begin{equation}
\label{eq:poisson2D_integral_discrete}
\displaystyle
\Phi(x_0, y_0) = 
-\sum\limits_{i=1}^{N_x-1}\sum\limits_{j=1}^{N_y-1}
\frac{q_{i,j}}{\sqrt{(x_i-x_0)^2+(y_i-y_0)^2}},
\end{equation}
where $q_{i,j} = G \Sigma_{i,j} h_x h_y$.

Direct calculation of the potential $\Phi(x_0, y_0)$ for all $(x_0, y_0)$ takes 
 $O(N_x^2 N_y^2)$ arithmetic operations. However, using the convolution theorem and Fast Fourier Transform the amount of operations can be reduced to $O(N_x N_y (\log N_x + \log N_y))$:
\begin{equation}
\label{eq:theorem_convolution}
\begin{array}{l}
\displaystyle
FFT[\Phi](\mathbf{k}) = -FFT[\rho](\mathbf{k})\cdot FFT[K](\mathbf{k}),\\
\Phi = -FFT^{-1}\left[ FFT[\rho] \cdot FFT[K] \right]
\end{array}
\end{equation}
where $FFT[\dots]$ is a two-dimensional Fast Fourier Transform, and $K$ is a kernel function that 
we defined as
$$
\displaystyle
K(x,y)=\left\{
\displaystyle
\begin{array}{rl}
& \frac{1}{0.5 \min (h_x, h_y)}, \ \ \sqrt{x^2+y^2} = 0,\\
& \frac{1}{\sqrt{x^2+y^2}}, \ \ \sqrt{x^2+y^2} > 0.
\end{array}
\right.
$$
To implement this algorithm we used the FFTW~\citep{fftw} library. 

\subsection{Coupling of SPH and grid-based Poisson equation solver}
\label{sec:coupling}

To calculate the gravitational forces acting from ensemble gravitational potential, first masses of individual SPH particles should be interpolated into density defined on a Cartesian grid, and then, when the value of potential is defined on the grid, forces should be calculated in nodes and inverse interpolation of mesh force into particles should be done.

There is some freedom in choosing the way to interpolate density and force.
E.g. \citet{Hydra} choose the triangular shaped cloud as an assignment function, and the 10-point differential operator for force calculation. In Gadget-2 \citep{Gadget2}, the Cloud-in-the-Cell assignment function \citep{Hockney} and then the 4-point differential operator are used. 

\begin{figure*}
  \includegraphics[width=0.8 \textwidth]{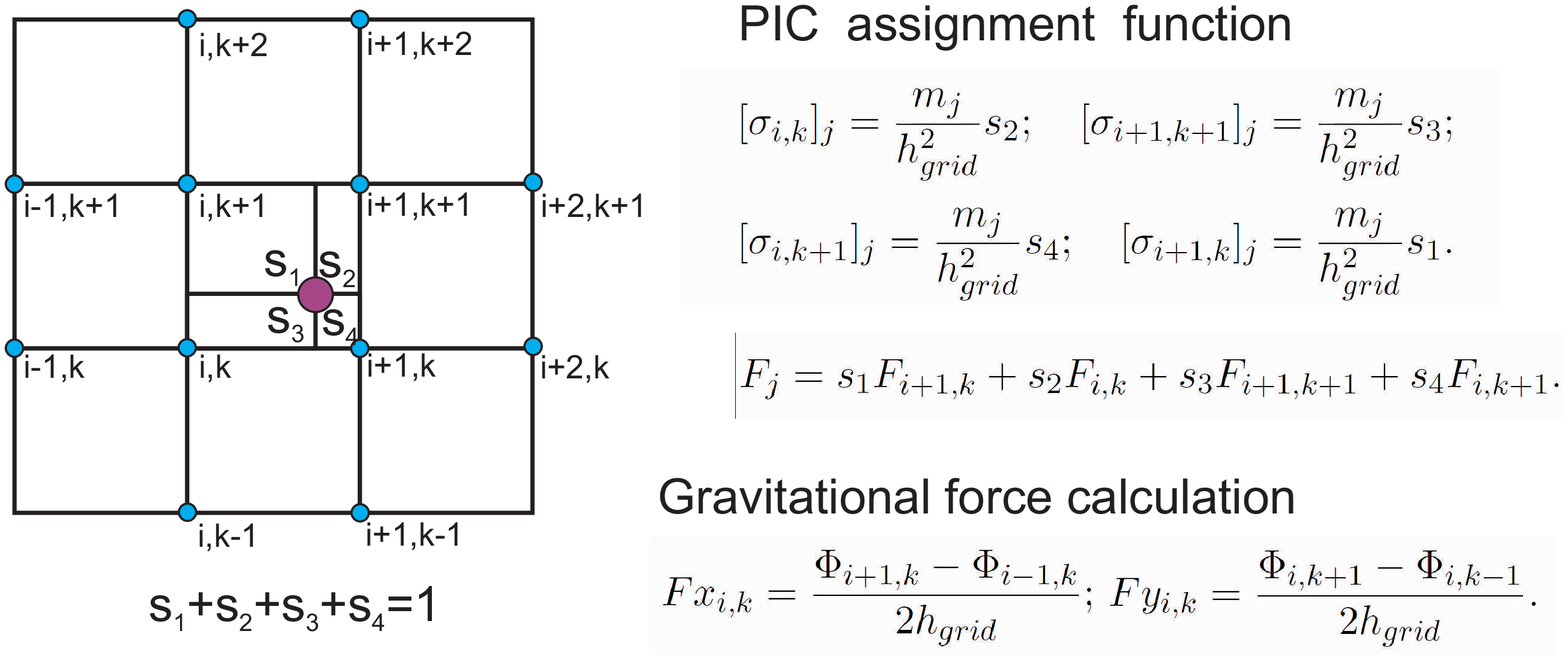} 
  \caption{Coupling of SPH particles to the grid: PIC assignment function and pattern to force calculation. Here $s_1$, $s_2$, $s_3$, and $s_4$ are areas of rectangular regions marked on the figure divided on the area of the whole cell.}
\label{fig:PIC}
\end{figure*}

In our implementation we used the Particle-in-the-Cell assignment function to construct density field on the mesh and interpolate force to a particle. Central difference (2-point differential operator) is used to calculate the forces in the mesh grid (see Fig.~\ref{fig:PIC}). 

Below we demonstrate the role of differential operator in the potential to force calculation.

To minimize the number of arithmetic operations in force calculation one can use forward difference for the left mesh cells and backward difference for the right mesh cells (in terms of Fig.\ref{fig:PIC}): $\displaystyle Fx_{i,k}=\frac{\Phi_{i+1,k}-\Phi_{i,k}}{h_{grid}}=Fx_{i+1,k}, \displaystyle Fy_{i,k}=\frac{\Phi_{i,k+1}-\Phi_{i,k}}{h_{grid}}=Fy_{i,k+1}$. For 2D case, this approach requires treating only four nearest mesh nodes where potential is defined to calculate the ensemble gravitational force
$$[Fx]_j=\frac{\Phi_{i+1,k}-\Phi_{i,k}}{h_{grid}}(s_1+s_2)+\frac{\Phi_{i+1,k+1}-\Phi_{i,k+1}}{h_{grid}}(s_3+s_4);$$
$$[Fy]_j=\frac{\Phi_{i,k+1}-\Phi_{i,k}}{h_{grid}}(s_2+s_4)+\frac{\Phi_{i+1,k+1}-\Phi_{i+1,k}}{h_{grid}}(s_1+s_3);$$
while application of central differences involves 12 nearest mesh nodes and more operations. Thus it makes the forward-backward approach suitable for calculation of the long-term disc evolution and structures without areas of high density. In case this scheme is applied to simulation of clump dynamics, a crude numerical artefact such as 'sticking clump' appears when the density peak reaches the threshold. 

Another important issue of coupling SPH with a grid-based solver for self-gravity is an acceptable relation between the length of discretization for gas dynamics properties $h_{SPH}$ and the potential gradient $h_{grid}$ calculation. \citet{SPHres} and \citet{Nelson} showed that a severe artificial imbalance between pressure and gradient forces can develop if a different length scale is applied. To exclude any enhancement of suppression of fragmentation caused by differences in hydrodynamical smoothing length and gravitational softening length, in this paper we used constant smoothing length $h_{SPH}=h_{grid}$. 

\section{Test results. Is it acceptable to treat more neighbors in SPH simulations?}

\label{sec:tests}

Applying SPH method to simulation of protostellar accretion from gravitationally unstable discs prone to fragmentation can be challenging because massive and dense fragments that form in the disc are also the areas of increased concentration of model particles. If a constant smoothing length is adopted, then it may result in an increased number of neighbours for each particle, which (1) increases the computational costs  for every time step and (2) may facilitate the development of the clumping or pairing instability. 

\begin{figure}
  \includegraphics[width=\columnwidth]{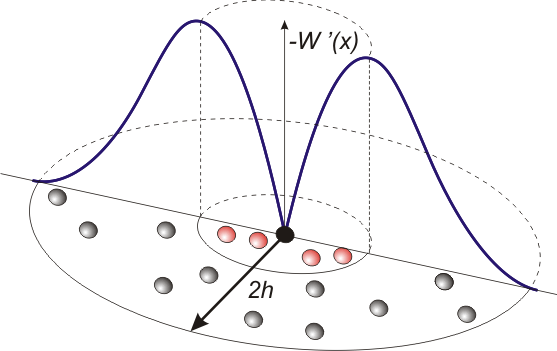} 
  \caption{Simplified essence of pairing instability. The shape of the first derivative of the cubic spline kernel used for calculation of forces in SPH. The repulsive force between black and grey particles behaves correctly (increases with decreasing the distance between particles), while between black and red particles behaves 'unphysically' (decreases with decreasing the distance between particles), which may result in the merging of several particles into one. }
\label{fig:Clumping}
\end{figure}

The simplified essence of the latter phenomenon in gas dynamics simulations was earlier described by \citet{Pairing}. They demonstrated that for closely spaced particles the repulsive force is underestimated in the case of a smoothing kernel has an inflection point. Inflection point of a kernel is a point where the first derivative of it has a maximum that does not coincide with the coordinate origin. In this case, as it can be seen from Fig.\ref{fig:Clumping}, decreasing the distance between two SPH particles results first in the pressure gradient reaching a local maximum and then decreasing as the two particles approach each other, while in reality decreasing the distance between two gas parcels results in a monotonous increase of the pressure gradient. Due to the underestimation of the repulsive force between closely spaced particles, the particles start moving towards each other. This convergence can last until their coordinates are merged into a single point. Since the accuracy of interpolation in SPH depends on regularity of particle distribution, merging of particles means the loss of approximation.  

Despite the pairing instability was earlier described as a result of diminution of the repulsive force between near-neighbour particles, now it is well understood that it can be caused by several sources. On the other hand, underestimation of the pressure gradient between closely spaced particles may lead not only to the loss of accuracy, but also facilitate the development of Jeans gravitational instability, which is a result of the competition between pressure gradient (repulsive force) and self-gravity of the volume (attractive force). For this reason we considered these two phenomenon as independent and on the test problem measured the effects of them on the obtained solution. More specifically, in Section~ \ref{sec:pressure} we estimate weather near-neighbour force diminution leads to pressure gradient underestimation that facilitates clump formation, when in Section~\ref{sec:kernel} we estimate the actual loss of interpolation nodes in our simulations of fragmenting protoplanetary disc. 

As a test problem we chose gaseous disc of 0.4~Solar Mass, and the central body mass equal to 0.8 Solar Mass. The disc extended from 10 to 100~au. The temperature varied from 90~K (10~au) to 30~K (100~au). Such configuration provides the initial value of Toomre parameter $Q<1$ for $R>50$~au, and the local Jeans length inside the interval 6 - 15~au. For such a disc, overdensity clump formation are expected after one orbital time of the disc periphery. 

For gravitational force calculation, the ensemble potential $\Phi$ was calculated on a regular Cartesian grid with the length of mesh cell $\displaystyle h_{car}=\frac{R_{disc}}{256}=0.39$~au. We used $1.6*10^5$ SPH particles to simulate the disc dynamics. \textbf{The time step was taken equal to 0.03~year and kept fixed in space and time during the simulation. This value guarantees that (1) orbit of individual particle rotating the protostar with Keplerian velocity at the inner edge of the disc is resolved at least by 1000 steps, (2) Courant number is less than 0.2 for all parts of the disc.} At the moment when disc fragmentation started, the minimum value of the local Jeans length reached 3~au and thus adopted resolution was enough to capture the disc fragmentation correctly (according to the criteria earlier discussed in detail by \citet{Truelove}, \citet{SPHres}, and \citet{Nelson} etc). 

It was underlined by \citet{Wendland} and we confirmed it from our practice that computational costs rise sublinearly with increasing the number of neighbours. In our test simulations actual CPU time for runs when we treat more than 1000 neighbours for some SPH particles is only 3-fold higher than with 20-30 neighbours. 

\subsection{Influence of different kernels on enhancement or suppression of fragmentation} 

\label{sec:pressure}

In this section we measure the effect of near-neighbour force diminution, an attribute of continuously differentiable kernels, on the dynamical outcome of fragmenting disc simulations. To do this we compare the results of the disc simulation with the same physical and numerical setups that differ only by the adopted kernels. First kernel was a classical cubic spline $W$ (\ref{eq:CubicSpline}) with its precise derivative $W'$ for force approximation. Second kernel was a kernel used by \citet{TC1992}(TC), where exact derivative of cubic spline $W'$ was substituted by corrected derivative $W'_*$ (\ref{eq:ThomasCouchman}) to ensure that the artificial phenomenon such as decreasing of the pressure gradient with decreasing of distance between particles is totally absent. This simple modification guarantees that simulations are free of near-neighbour force diminution at the cost of inconsistency of kernel normalization. The third kernel is quintic Wendland function kernel (\ref{eq:WendlandKernel}). For standard cubic spline kernel, the distance that provides a repulsive force underestimation between particles is smaller than two thirds of the smoothing length, while for fifth order Wendland polynomial this distance is about half of the smoothing length.   

\begin{figure}
  \includegraphics[width=\columnwidth]{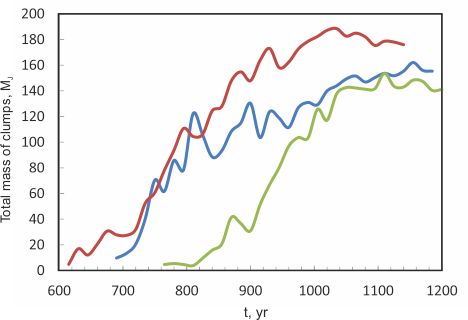} 
  \caption{Total mass of clumps obtained in simulation of 0.4 Solar mass disc with different kernels applied: Thomas-Couchman (green line), cubic spline (blue line) and Wendland function (red line).}
\label{fig:Kernel Comparation}
\end{figure}

In the obtained results we found that the whole dynamical picture is very similar for all kernels: first fragments appeared in the outer part of the disc, and after hundred of years fragmentation of inner part of the disc took place. \textbf{In simulations with  all kernels the mass of appeared clump varied from 2 to 10 $M_{J}$. The mass of clump depends on the radius of its formation and varies drastically during its migration in the disc (see eg. Fig.(\ref{fig:ClumpTemp})) due to accretion of gas onto the clump and matter demolition due to tidal forces.} Fig.\ref{fig:Kernel Comparation} demonstrates the total mass of clumps obtained in simulations with different kernels. Total mass of clumps was found as a sum of all clumps detected in the disc using HOP algorithm~\citep{HOP}, while the mass of individual clump was calculated as a mass of all particles with density higher than 0.5 maximum density in the clump. Evidently, in all simulations the total mass of clumps behaves as S-type curve, typical for instability development. More specifically, during couple of hundred of years after the first fragment appeared in the disc, the mass of clump grows near linearly and than reaches its plateau value. 

One can found that the slope of Wendland curve is very similar to the slope of TC kernel, despite the fact that TC curve is shifted for 200~yr. This shifting is a result of construction of TC kernel, which is free of near-neighbour force diminution, featured by Cubic and Wendland kernels. Moreover, later appearance of fragment prooves, that origin of initial distortions depends on adopted kernel, but the development of the instability is reproduced in a similar way by all kernels and thus independent on the SPH implementation.  

One can see also that the Cubic spline curves on the stage of intense growth of total mass of clumps almost coincides with Wendland curve. But due to the fact that dynamics of multiple gravitating objects in the disc becomes stochastic, e.g. \citep{Boss}, we obtained the differences in the total mass of clumps in simulations with different kernels. E.g. the rapid decrease of total mass of clumps at 800~yr in Cubic spline simulation is a result of clumps accretion onto the adsorbing cell, caused by special arrangement of clumps in the disc stochastically formed in this run and not repeated in simulation with Wendland kernel.  

\subsection{Particle noise for large number of neighbours - kernel comparison}

\label{sec:kernel}

In this section we will focus on another aspect of the long-term calculation of fragmenting discs with SPH - keeping the regularity of particle distribution or increasing of particle noise. 

\citet{Wendland} recommended Wendland function as a kernel providing better convergence at a higher number of neighbours. To see if this kernel gives a benefit in keeping the particle order for gas dynamics with artificial viscosity simulations of clump formation, we compare the results obtained with classical cubic spline, TC kernel and quintic Wendland function. 

\begin{figure}
 \includegraphics[width=\columnwidth]{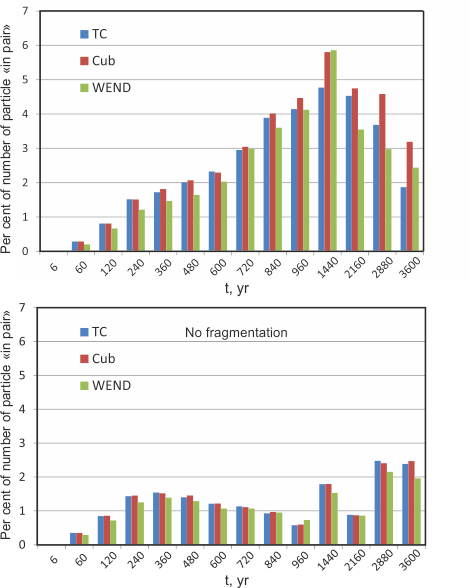} 
  \caption{Number of "particle in pair" (particles for which order koefficient $q_i$ is less than 0.01) in per cent of total ammount of SPH particles at different time instances for three adopted kernels: Thomas-Couchman (blue), Cubic spline (red), Wendland quintic polynom (green). Top panel - simulation of fragmenting disc of 0.4 Solar mass, bottom panel - simulation of nonfragmenting disc of 0.05 Solar mass.}
\label{fig:pairs}
\end{figure}

We simulated the dynamics of the disc of 0.4 Solar Mass described in the previous section applying these three kernels. To estimate the value of particle disorder in the obtained results, for every particle we calculate a value $q_i=\displaystyle\frac{r_{min,i}}{h_{SPH}}\sqrt{\frac{\sqrt 3 N_{neib,i}}{2 \pi}}$ - the order coefficient similar to the indicator of regularity of particle distribution used by \citet{Wendland}. Here, $r_{min,i}$ is the distance between particle $i$ and its closest neighbour, $h_{SPH}$ - the hydrodynamical smoothing length of SPH particles, number $\displaystyle\sqrt{\frac{2 \pi}{\sqrt 3 N_{neib,i}}}$ estimates the ratio of particle spacing to smoothing length for the case of uniform distribution of $N_{neib,i}$ particles inside the circle of radius $2 h_{SPH}$ on a triangular lattice grid (which provides the tightest regular packing).

Then we calculate the number of particles 'in pair', for which $q_i$ is less than $0.01$. The results can be seen on the top panel of Fig.\ref{fig:pairs}. The first column of Fig.\ref{fig:pairs} shows that in the beginning of the simulations with all kernels particles are ordered because number of pairs is almost zero. With the development of the instability the number of pairs grow near linearly with the time for all kernels. By the time 1440~yr (multiple fragments were formed in the disc simulated with all kernels) about 6 per cent of total number of particles are in pair, which means that we keep about 94 per cent of interpolation nodes. After reaching the peak, the number of particle 'in pair' decreases monotonously from 6 to 2-3 per cent of initial total amount of particles. 

Evidently that TC kernel that guaranties absence of 'unphysical' pressure gradient underestimation provides almost the same results as cubic spline during the process of clump formation. It confirms the already understood idea that pairing due to near-neighbour force diminution is not the only reason of particle noise appearance. 

Bottom panel of Fig.\ref{fig:pairs} demonstrates the number of particle 'in pair' for the disc of 0.05 Solar mass without any fragments. Visual examination of Fig.\ref{fig:Kernel Comparation} and Fig.\ref{fig:pairs} indicates that growth of total mass of clumps is accompanied by increasing of number of particle 'in pair' from 2 to 6 per cent. On the contrary, in simulation of low-massive discs, during the period between 360 and 1440 yr, monotonous decreasing of 'disordered particle' took place. By comparison of top and bottom panels, one can conclude that, on average, dynamical processes in the discs have stronger influence on particle disorder than type of implemented kernel. 

One can see that Wendland function demonstrates systematic benefit in keeping particle order. This benefit is significant in the very begining of simulations (60 yr), when application of Wendland kernel allows decreasing the number of pair about on 30 per cent comparing to results with cubic spline and Thomas-Couchman kernel and in the stage, when overdensity clumps are already formed (later than 2160 yr). We adopted Wendland kernel for our further simulations due to the benefit in keeping particle order for large number of neighbours.

\section{Disc gravitational fragmentation: numerical resolution study}
\label{sec:ResGeneral}

In this section, we perform several test simulations of disc dynamics in order to determine the dependence of our results on the numerical resolution. The resolution requirement for the correct simulation of self-gravitating discs was formulated in terms of the Jeans mass or, equivalently, the Jeans length by \citet{Truelove}. In addition, \citet{SPHres} and  \citet{Nelson} demonstrated that for the particle-based simulations the relation between the gravitational softening length and the hydrodynamical smoothing length is another resolution requirement that needs to be taken together with the Jeans length requirement. More specifically, the Jeans mass needs to be resolved by at least 10-12 SPH particles and the gravitational softening length needs to be equal to the hydrodynamical smoothing length. The second requirement has to be taken into account to avoid a numerical imbalance between pressure and gravitational forces. Moreover, for two-dimensional and three-dimensional calculations of disc dynamics an additional parameter, the disc scale-height, should be resolved properly (at least by several hydrodynamical smoothing lengths in the equatorial plane) as discussed in detail by, e.g., \citet{LodatoClarke}.

Our numerical model - a grid-based solver for the Poisson equation in combination with the SPH - implies that we have fixed in space and time the gravitational softening length, which now becomes equal to the size of the corresponding grid cell of our numerical grid for the Poisson solver. The hydrodynamical smoothing length is set equal to the gravitational softening length to fulfill the abovementioned requirement. 

The initial configuration of the disc is described in sec.\ref{sec:init} and is given in Fig.\ref{fig:init}. The standard numerical resolution of the disc is 160000 SPH particles and the increased resolution is 640000 SPH particles. The computational domain has a size of $400\times 400$~au$^2$. Computations with 160 000 particles were done on $1024\times1024$ grid cells and the computations with 640 000 particles were done on $2048\times2048$ grid cells.

The list of models is given in Table 1. For each model, the number indicates the mass of the disc, the letter indicates the adopted number of particles (S - standard, I - increased). If disc fragmentation takes place, the name of the model is marked in bold. For the discs without fragmentation the integration time is 6000yr, while calculations with fragmentation are done for as long as possible. 

\begin{figure}
  \includegraphics[width=\columnwidth]{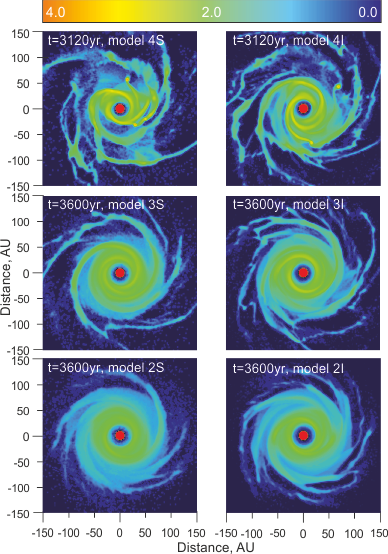} 
  \caption{The logarithm of gas surface density in different time moments obtained with different numbers of SPH particles: first column - 160 000, second column - 640 000. Top line - the mass of the disc is 0.25 Solar mass, middle line - the mass of the disc is 0.2 Solar mass, bottom line - the mass of the disc is 0.15 Solar mass.}
\label{fig:converg2}
\end{figure}

Fig.\ref{fig:converg2} presents the gas surface density distribution; the top, middle and bottom panels correspond to disc masses of $M_{\rm d}=0.25~M_\odot$, 0.2~$M_\odot$, and 0.15~$M_\odot$, respectively. For all simulations the mass of the protostar is equal to $0.8~M_\odot$. In particular, the left and right columns provide the results for 160000 and 640000 SPH particles with the corresponding smoothing lengths $0.39$~au and $0.195$~au. Evidently, in all models the outcome does not depend on the adopted resolution: the most massive model $M_{\rm d}=0.25~M_\odot$ shows disc fragmentation, while the least massive does not. This behavior is expected from the radial distribution of the Toomre $Q$-parameter shown in Fig.\ref{fig:init}. The $M_{\rm d}=0.15~M_\odot$ model has a $Q$-parameter greater than 1.0, a fiducial critical value for disc fragmentation, almost throughout the whole disc extent, while the $M_{\rm d}=0.25~M_\odot$ model has $Q<1.0$ for $r\ge60$~au, meaning that nearly half of the initial disc extent is prone to gravitational fragmentation. We checked the disc behavior for models with disc masses of $0.3~M_\odot$ and $0.1~M_\odot$ and confirmed the revealed tendency: discs with mass $\ge 0.25~M_\odot$ fragment regardless of the number of SPH particles, whereas discs with mass $\le 0.2~M_\odot$ do not. 

\begin{figure}
  \includegraphics[width=\columnwidth]{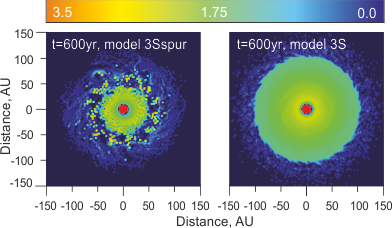} 
  \caption{: Logarithm of gas surface density obtained in 600 year of disc dynamics. The mass of the disc is 0.2 Solar mass. All calculations are done with 160 000 SPH particles. Left panel - constant smoothing length is used,  $\displaystyle \frac{h_{SPH}}{h_{grid}}=3$. For this calculation minimum local Jeans length is 6au, when linear size of clusters is about 1au. Right panel - constant smoothing length is used, $\displaystyle \frac{h_{SPH}}{h_{grid}}=1$ .}
\label{fig:artClumps}
\end{figure}

As the next step, we demonstrate that the condition $h_{SPH}>h_{grid}$ can lead to overestimation of the calculated gravity force as compared to the pressure force. Several authors, e.g., \citet{Durisen, Nelson} demonstrated that such an imbalance in forces has a strong affect on the dynamical outcome of disc simulations. Fig.~\ref{fig:artClumps} presents the gas surface density distribution the $M_{\rm d}=0.2~M_\odot$ model after 600 yr of disc evolution obtained using different numerical setups: the left panel corresponds to simulations with a constant smoothing length $h_{SPH}=3~h_{grid}$ and the right panel corresponds to model~3S with a constant smoothing length  $h_{SPH}=h_{grid}$. One can see that the model with an increased hydrodynamical smoothing length produces multiple artificial clumping in the disc which are absent in model 3S. It is important to note that the minimum value of the local Jeans length for the modelled disc configuration is 6~au, which is adequately resolved by the adopted $h_{SPH}=0.39$~au and $h_{SPH}=1.2$~au. At the same time, the linear size of obtained clumps is about 1~au, much smaller than the Jeans length, meaning that fragmentation in this model is indeed spurious. These results are in agreement both with theoretical expectations and with numerical examples of artificial clumping found by \citet{Nelson}.

Table 1 demonstrates that except for model 3Sspur with the numerical setup specially designed to produce artificial clumps, the dynamical outcome of disc evolution is independent from the numerical resolution and in agreement with the initial distribution of the Toomre parameter $Q$.

We should note that discussion on necessary and sufficient criteria of disc fragmentation is a separate stream of computational aspects of protoplanet formation simulation and is beyond the scope of our paper. Due to this fact we cite only limited number of papers on this problem. After the work by \citet{Gammie}, efforts of several groups were directed to evaluate sharp boundary of disc fragmentation from numerical simulation: e.g.\citet{MeruBate,LodatoClarke,RiceCool,Paardekooper,YoungClarke} etc. Some systematic effects of the applied particle or grid-based method \citep{MeruBate}, resolution of the Jeans length and mass, Toomre length and mass, \citep{Nelson}, disc height \citep{LodatoClarke,YoungClarke}, form of artificial viscosity and its parameters, way of cooling implementation \citep{RiceCool}, method of coupling gas dynamics and gravity solver \citep{SPHres} are described as an elements responsible for fragmentation or absence of fragmentation. \citet{Paardekooper} presented analytical arguments supporting the idea that fragmentation is a stochastic process. \citet{Takahashi} demonstrated that spiral arms fragment only when $Q<0.6$, using numerical simulations and linear stability analysis for the self-gravitating spiral arms. \citet{Stoyanovskaya2016CMP} demonstrated that that the process of clump formation can be characterized by an average growth rate of the total mass of fragments in the disc; this rate is strongly dependent on the physical parameters of the disc and is slightly dependent on the parameters of the numerical model.  \citet{SnytnikovStoyanovskaya2016CMP} substantiated the mathematical correctness of numerical models based on the SPH for simulations of gravitational instability development in circumstellar discs.   

In the next section, we apply the developed numerical model to protostellar accretion simulation. The influence of the numerical resolution on the obtained accretion rate will be estimated in Section \ref{sec:resolutionAcrate}. 

\begin{table*}
 \begin{minipage}{140mm}
  \caption{List of computational experiments. If any fragment appear in the disc, the name of the run is marked in bold. The mass of the protostar is equal to 0.8 Solar Mass. }
  \begin{tabular}{@{}llllll@{}}
  \hline
   Mass of the Disc (Solar Mass)     & 0.1 &  0.15 &  0.2  &  0.25 &  0.3 \\
 
\hline

 $1024 \times 1024$ cells, 160 000 SPH, constant smoothing length $h_{SPH}=h_{grid}$& 1S &  2S &  3S  &  \textbf{4S} &  \textbf{5S}  \\
 
\hline 
 
 $1024 \times 1024$ cells, 160 000 SPH, constant smoothing length $h_{SPH}=3h_{grid}$&   &    &  \textbf{3SSpur}  &   &    \\
  
\hline

 $2048 \times 2048$ cells, 640 000 SPH, constant smoothing length $h_{SPH}=h_{grid}$& 1I &  2I &  3I  &  \textbf{4I} &  \textbf{5I}  \\
  
\hline

\end{tabular}
\end{minipage}
\end{table*}    

\section{Results of protostellar accretion simulations}
\label{sec:results}
 
To compare the accretion rates in self-gravitating discs with and without fragmentation, we simulated the dynamics of the disc extended from 10 to 100~au around a protostar with mass 0.8~$M_\odot$. We took the initial configuration of the disc similar to that formed in numerical hydrodynamics simulations of cloud core collapse by \citet{Vorobyov2010}. The initial temperature declines from 90~K at 10~au to 30~K at 100~au, being inversely proportional to the square root of the radial distance from the star. Discs with masses from 0.1 to 0.3~$M_\odot$ were considered with the surface density inversely proportional to the radial distance. 

The computational domain has a size of $400\times 400$~au$^2$. The standard numerical resolution of the disc is 160000 SPH particles and $1024\times1024$ grid cells, the increased resolution is 640000 SPH particles and $2048\times2048$ grid cells. \textbf{As in section \ref{sec:tests}, we take the time step equal to 0.03~yr and keep it fixed in time and space during the simulations with standard and increased resolution. This time step guarantees the Courant number less than 0.4 for all parts of the disc for models with increased resolution.}  
 
The criterion for accretion of SPH particles depends on the distance from the protostar: particles approaching the protostar closer than $R_{\rm cell}$ are considered as accreted and transfer their mass onto the protostar. Since the gas flow around the inner sink cell is supersonic in the azimuthal direction but is usually subsonic in the radial one, the accurate treatment of the inner boundary requires the development of special schemes which take into account a smooth transition of hydrodynamic variables and gravitational potential through the inner boundary. We do not employ them in the present study, but note that this may lead to an artificial depression in the gas density near the sink cell. We plan to work on this artefact in a future study. To avoid too small time steps, we set the radius of the sink cell equal to $R_{\rm cell}=10$~au.

In this section, we describe different dynamical outcomes of our numerical simulations, focusing particularly on models in which the episodic character of protostellar accretion reveals itself. We define episodic bursts as sharp increases in the mass accretion rate that are greater than the quiescent accretion (immediately preceding the burst) by at least 1.5 orders of magnitude. For lower variations or oscillations in the accretion rate the term ''variable accretion'' is reserved. In subsection \ref{sec:resolutionAcrate} we provide results of our resolution study. In subsection \ref{sec:previous}, we describe episodic accretion bursts associated with destruction of infalling clumps. In subsection \ref{sec:new} we describe accretion bursts caused by perturbations of the inner part of the disc.

The mass accretion rate is calculated from the protostar mass $M_{\rm c}$ using the following expression: 
\begin{equation}
\label{eq:Acrate}
\dot{M}=\displaystyle\frac{M_c(t+\tau_{acc})-M_c(t)}{\tau_{acc}}.
\end{equation}
\textbf{Thanks to our numerical method the minimal portion of mass accreted onto the protostar is equal to the mass of individual SPH particle. Thus to avoid oscillations in calculated mass accretion rate caused by numerical resolution, we must use 
\begin{equation}
\label{eq:tauAc}
\tau_{acc}>\displaystyle\frac{m_{SPH}}{\dot{M}}=\tau_{min}.
\end{equation}
Eq.(\ref{eq:tauAc}) (1) shows that we can not study with our model perturbation in accretion rate shorter then $\tau_{min}$, (2) provides necessary, but not sufficient condition to obtain smooth accretion rate. In our simulations with standard resolution $m_{SPH}$ varied from $6.25 \times 10^{-7}M_{\odot}$ to $18.75 \times 10^{-7}M_{\odot}$ and with increased resolution --- from $1.56 \times 10^{-7}M_{\odot}$ to $4.69 \times 10^{-7}M_{\odot}$. On the other hand, the typical accretion rate for many observable discs is rather slow, about $\dot{M}=10^{-9}-10^{-6}$ $M_{\odot}$~yr$^{-1}$. Thus according to Eq.(\ref{eq:tauAc}) we used $\tau_{acc}$ much greater then adopted time step $0.03$~yr$^{-1}$. In particular $\tau_{acc}$ was varied from 3~yr (when we focus on bursts resolution) to 120~yr (when we focus on tendencies in typical accretion rate). The influence of $\tau_{acc}$ variation is demonstrated in \ref{sec:TauAcc}.} The exact value of $\tau_{acc}$ is provided in the figure captions. 
 
\subsection{Mass accretion rate in fragmenting and non-fragmenting discs: numerical resolution study}

\label{sec:resolutionAcrate}

\begin{figure}
  \includegraphics[width=\columnwidth]{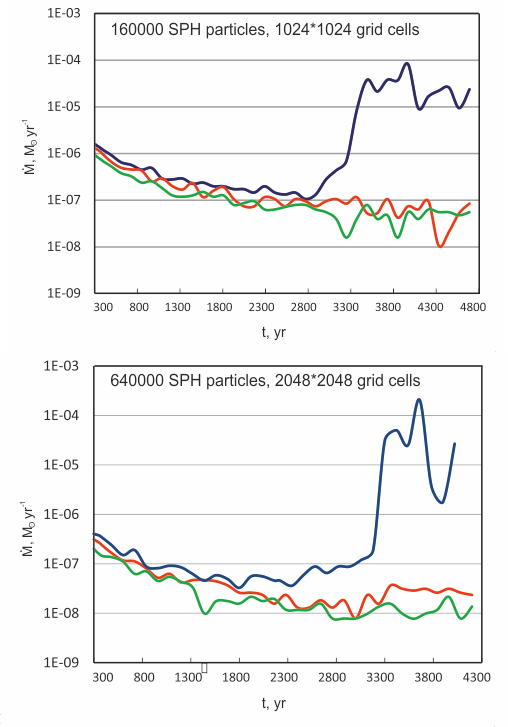} 
  \caption{Accretion rates for models with disc masses of 0.15 (green line), 0.2 (red line) and 0.25 (blue line) Solar mass. Top figure - models with standard resolution 2S, 3S, 4S. Bottom figure - models with increased resolution 2I, 3I, 4I. For top and bottom panel the time step to calculate the accretion rate $\tau_{acc}$ is equal to 120~yr (see \ref{sec:TauAcc} for rationale of $\tau_{acc}$ choice).}
\label{fig:AcRate}
\end{figure}
 
fAs a first step, we check if our model can reproduce qualitatively different accretion histories as expected for fragmenting and non-fragmenting discs by comparing $\dot{M}$ for models with disc masses of 0.15, 0.2, and 0.25~$M_\odot$. The top panel in Fig.~\ref{fig:AcRate} presents $\dot{M}$ vs. time calculated for models 2S, 3S, and 4S for one orbital period of an outer disc. \textbf{Since we focus on general tendencies in accretion rate we use $\tau_{acc}=120$~yr.} As expected, the non-fragmenting models 2S and 3S exhibit rather smooth accretion rates in the  $(10^{-8} - 10^{-6})~M_\odot$~yr$^{-1}$ range. A rather steep decline of $\dot{M}$ with time was most likely caused by the development of spiral modes, which first drove the system out of equilibrium and then brought it towards a new steady state configuration. The greater the mass of the disc is, the higher accretion rate it provides: increasing the mass of the disc from 0.15 to 0.2~ $M_\odot$ the accretion rate becomes a factor of 1.3 higher, which is in agreement with a near-linear correlation between the disc accretion rate and the disc mass found previously by, e.g., \citet{VorobyovBasu2008}. 

On the other hand, the fragmenting model~4S with disc mass 0.25~$M_\odot$ demonstrates the development of variable accretion with episodic bursts after 3300 years of its evolution. These results are in agreement with numerical simulations of, e.g., \citet{VorobyovBasu2006,VorobyovBasu2010} and \citet{MachidaAc}, who found that the burst mode of accretion develops in self-gravitating discs prone to fragmentation, in which fragments are driven onto the star due to the loss of angular momentum via gravitational interaction with spiral arms and other fragments. 

The bottom panel in Fig.~\ref{fig:AcRate} presents $\dot{M}$ vs. time calculated for  the same models as in the top panel, but with an increased numerical resolution. Evidently, the most essential features of the mass accretion rate are captured in all groups of models and are independent of numerical resolution. Discs without fragments feature decreasing accretion rates in the interval from $10^{-6}~M_\odot$~yr$^{-1}$ to $10^{-8}~M_{\odot}$~yr$^{-1}$ with short-term variations less than one order of magnitude. Fragmenting discs show an accretion burst with the rate that is 2-3 orders of magnitude higher than in the non-fragmenting discs. All groups of models demonstrate also a nearly monotonous dependence of the accretion rate in the quiescent period on the mass of the disc. On average, more massive discs provide higher accretion rates in the quiescent phase. One can also see that the accretion rate in the quiescent phase is somewhat sensitive to the numerical resolution. Increasing the number of SPH-particles from 160 000 (top panel in Fig.\ref{fig:AcRate}) to 640 000 (bottom panel in Fig.\ref{fig:AcRate}) results in a decrease in the accretion rate by a factor of several during the first 2500 yr of disc evolution. These results indicate that viscous torques associated with numerical viscosity may affect somewhat the quiescent accretion rate and further convergence studies are needed to address this issue. \textbf{Since we focused on general tendencies in accretion rate, we used $\tau_{acc}=120$~yr for all models. This value was found to be sufficiently long to obtain smooth quiescent phase of accretion, but too long to reproduce duration and amplitude of bursts correctly.} 

\subsection{Clump destruction and associated episodic accretion bursts}
\label{sec:previous}

\begin{figure}
  \includegraphics[width=\columnwidth]{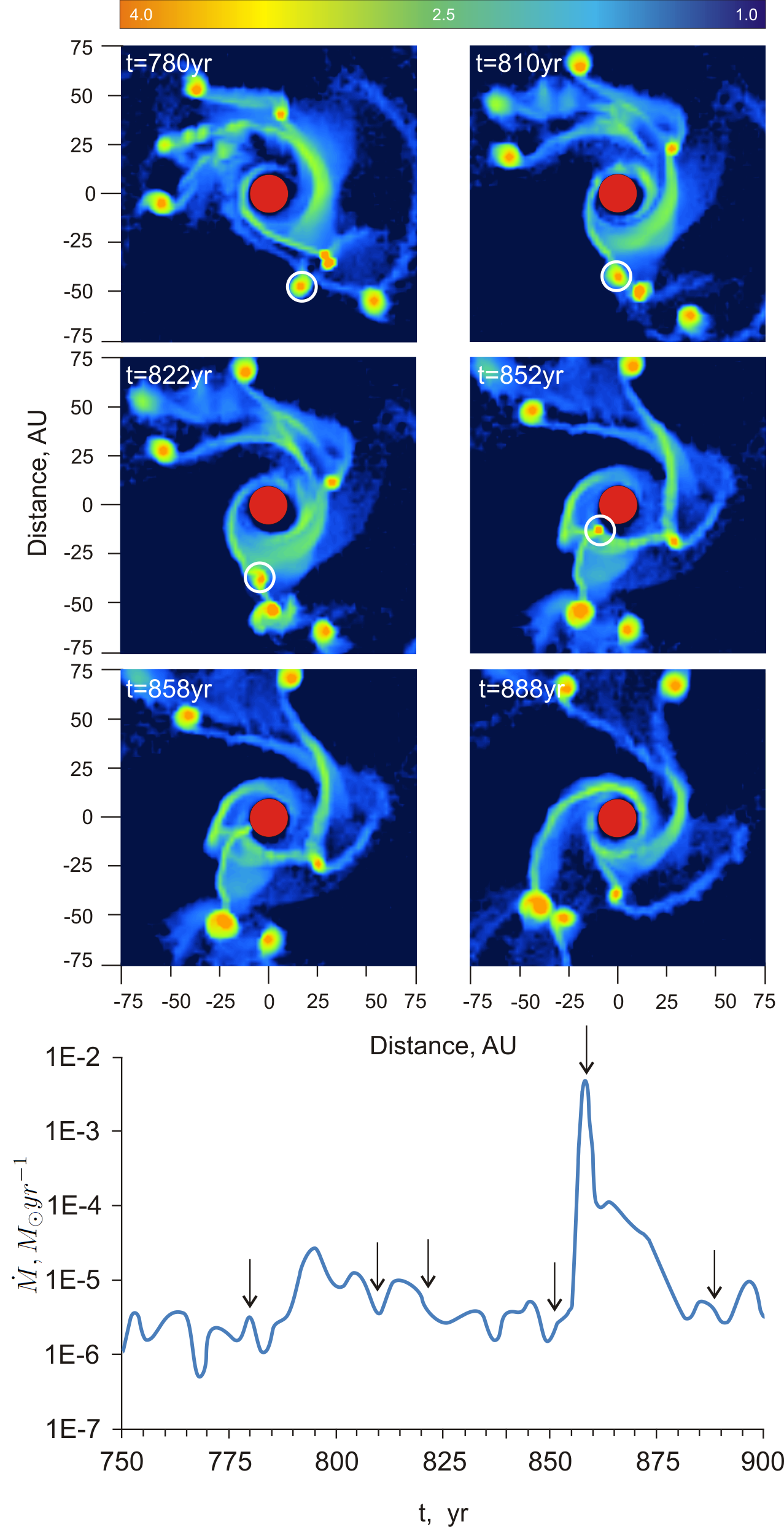} 
  \caption{Top - the logarithm of gas surface density obtained in 780, 810, 822, 852, 858, and 888 years of disc dynamics. The mass of the disc is 0.25 Solar mass (\textbf{model 5S}). Bottom - the accretion rate in terms of $M_{\odot} y^{-1}$ for the same period. Time step to calculate the accretion rate is equal to 3 year. Moments of snapshots are indicated with arrows on the accretion rate plot.}
  \label{fig:IsolatedBurst}
\end{figure}

Recent numerical simulations of gravitationally unstable discs by \citet{VorobyovBasu2015} predicted the existence of isolated burst, where the peaks in $\dot{M}$ are separated by prolonged periods (a few $\times10^3$~yr) of quiescent accretion. Isolated bursts are caused by accretion of dense and compact clumps that can keep their near-spherical shape when approaching the star. However, the global disc evolution models allow simulating the dynamics of these clumps only down to a distance of several au from the star, where an absorbing sink cell is usually introduced, thus neglecting all effects that can occur inside the adsorbing cell. Among them are the possible tidal destruction of the clump or further contraction, which may be accompanied by the clump-disc mass exchange. For this case, complementary models for processes inside the adsorbing cells were developed \citep[e.g.][]{NayakshinLodato} demonstrating that clump disruption due to tidal forces near the star can indeed produce accretion bursts. 

Figure~\ref{fig:IsolatedBurst} presents the disc evolution in model~5S, capturing an isolated accretion burst that is caused by the inward migration of one of the clumps highlighted by the white circles. The bottom panel shows the corresponding mass accretion rate through the inner sink cell. The circled clump is initially located at a distance of around 50~au, but starts migrating towards the star at $t=810$~yr owing to gravitational interaction with nearby trailing fragments that exert a negative torque on it. At $t=855$~yr, the circled clump passes through the adsorbing cell (10 au) keeping its original near-spherical shape and causing a strong accretion burst ($t=860$~yr). The mass of the clump is about $8~M_{\rm J}$ and the peak value of the mass accretion rate is equal to $5 \times 10^{-3} M_{\odot}$~yr$^{-1}$. During the next several decades the accretion rate comes back to its quiescent value. We note that other fragments continue moving on near circular orbits, indicating that the fast inward migration of clumps is a rather stochastic phenomenon which requires a certain arrangement of clumps.  
       
  \begin{figure}
  \includegraphics[width=\columnwidth]{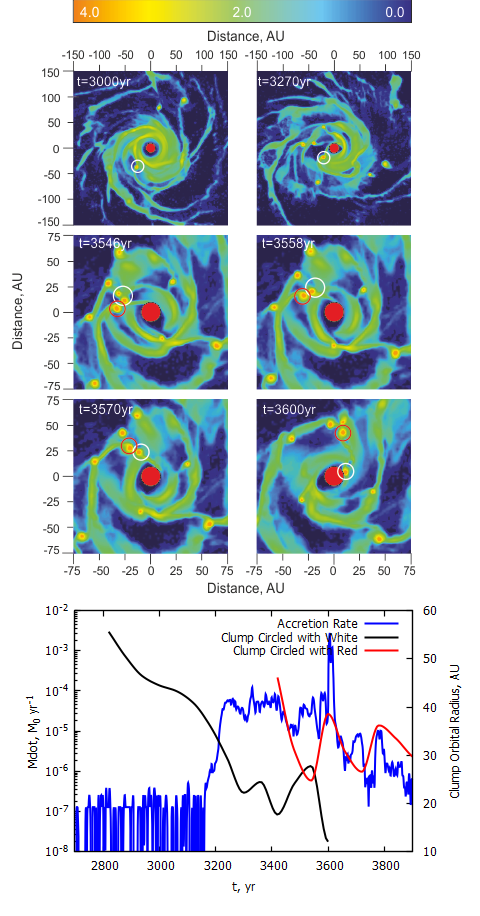} 
  \caption{Top: the logarithm of gas surface density obtained in 3000, 3270, 3546, 3558, 3570, and 3600 year of disc dynamics. The mass of the disc is 0.25 Solar mass (model \textbf{4I}). Bottom - the accretion rate in terms of $M_{\odot}~yr^{-1}$ for the same period (blue line), the orbital radius of the clumps circled with white (black line) and circled with red (red line). Time step to calculate the accretion rate is equal to 3~yr.}
  \label{fig:ClumpBurst}
   \end{figure}

\subsection{Clump triggered clustered burst mode}
\label{sec:new}

In this subsection, we present an example of accretion bursts caused gaseous clumps orbiting near the star. In contrast to the previously considered case, these clumps do not fall on to the protostar. This regime was previously described by \citet{MachidaAc}, who demonstrated that a planetary-sized object orbiting the protostar disturbs the inner part of the disc and promotes protostellar accretion. Fig.~\ref{fig:ClumpBurst} shows the gas surface density (in log~g~cm$^{-2}$) obtained in model 4I at six different evolutionary times. We note that the top row of disc images has a twice greater spatial extent than the middle and bottom rows. The fragment responsible for the accretion bursts is outlined by white circles. The bottom panels present the mass accretion rate through the central sink cell and the position of the highlighted clump. 

The mass accretion rate during the initial several hundred years of evolution gradually declines with time and exhibits a sharp increase from $10^{-7}~M_\odot$~yr$^{-1}$ to $10^{-4}~M_\odot$~yr$^{-1}$ at $t~\approx 3200$~yr. In the subsequent evolution, $\dot{M}$ stays at an elevated value, shows a strong peak at $t\approx 3600$~yr, and finally declines to a nearly pre-burst value after $t=3650$~yr.

A visual inspection of Fig.~\ref{fig:ClumpBurst} reveals that the highlighted clump (responsible for the burst) initially orbits the protostar at a distance of $\approx~60$~au. The mass of the clump is about $6~M_J$. The first increase in the mass accretion rate is concurrent with the time instance when the clump starts migrating inward and approaches a radial distance of $\approx 30$~au. However, unlike the previously considered case, the clump does not immediately cross the sink cell, but stops migrating inward and continues orbiting the protostar at a distance of about 15-30 au. At the time instance $t=3546$~yr, shown in the left middle panel of Fig.\ref{fig:ClumpBurst}, the clump mergers with another smaller clump of $3~M_J$. At the same time, a chance encounter of our clump with another massive clump outlined by the red circle causes the gravitational exchange of angular momentum between the two clumps. This results in the fast inward migration of one clump and the ejection to a more distant orbit of the other. Just after $t=3600$~yr, shown in the right bottom panel of Fig.~\ref{fig:ClumpBurst}, the closer clump crosses the adsorbing boundary and caused a strong accretion burst.

We compared the radial velocity of approaching clumps that indicates their presence in the inner part of the disc to the case where an isolated burst without prognostic modes was generated, and found that they differ by several fold. The clump that generates an isolated burst has the radial velocity about 1~au~yr$^{-1}$, while the clumps that generate clustered bursts have the value about $0.15-0.25$~au~yr$^{-1}$. \textbf{For more evidential comparison of accretion history produced by slowly and rapidly migrating clumps see \ref{sec:Petterns}.}   

The mechanism of clump-triggered burst mode requires further investigation, probably having the same roots as the triggered fragmentation of self-gravitating discs reported by citet{ArmitTrig,ClumpMigr,MeruTrig}. Further investigation of this scenario is necessary to constrain the properties of the clump and the disc, responsible for generation of triggered oscillating mode of accretion bursts.

\section{Conclusions} 
 
We applied a combination of Smoothed particle hydrodynamics with a grid-based solver of Poisson equation to simulation of mass accretion in massive gravitationally unstable discs. We found that this combination of methods allows treating the formation and dynamics of high-density clumps in massive gaseous discs with acceptable time-stepping. 

We confirmed that the dynamical processes in self-gravitating discs can produce the burst mode of accretion (wherein long periods of quiescent accretion are interspersed with short but intense accretion bursts) by means of disc fragmentation followed by the inward migration of the gaseous clumps on to the protostar. Besides the short-term bursts (10-40~yr) triggered by the rapid infall of the clump on to the protostar \citep{VorobyovBasu2010,VorobyovBasu2015}, our modeling predicts prolonged periods (200--300~yr) of elevated accretion culminated with a strong burst. The latter events are caused by the clumps tentatively halting their fast inward migration at distances $\sim~15-25$~au followed by rapid infall on to the protostar. A similar effect of the clump orbiting at $\sim 10$~au and triggering repetitive bursts was earlier reported by, e.g., \citet{MachidaAc}. In our case, however, the close-orbit clump sustains a high-rate accretion, $\sim (10^{-5}-10^{-4})~M_\odot$~yr$^{-1}$, for several hundreds of years and causes one strong burst when it ultimately falls on to the star.

It is known that a companion in an eccentric binary system, can trigger FU-Orionis-type accretion-luminosity bursts during the close approach with the primary \citep{BonnellBastien, Pfalzner2008}. In our case, an approaching clump may be regarded as such a disturber, albeit with a smaller mass. However, unlike the binary case, the clump can linger on a quasi-stable orbit near the protostar causing elevated accretion rates, while the companion would quickly recede to a larger distance. In this context, it is interesting to note that the FU Orionis itself, a binary system in the outburst for the last 80 years, has a companion at a projected distance of about 230 au \citep{Wang} and \citep{ReipurthAspin}. This is too far from the primary star to be consistent with the timing of the outburst, if the current burst indeed is caused by this companion. This led \citet{BeckAspin} to suggest that there might be another unseen companion in the inner disc regions of FU Orionis. In view of our numerical simulations, we suggest that this might be a massive gaseous clump formed via disc fragmentation, which might have been triggered by the past close encounter with the FU Orionis companion \citep{Thies2010}.

\section*{Acknowledgements}

OS was supported by Grant of President of Russian Federation MK 5915.2016.1, OS and NS was supported by RFBR grant 160700916. OS is grateful to the Centre for International Cooperation and Mobility of OeAD. OS, VS and NS  are grateful to the Ministry of Science and Education of the Russian Federation for a partial support of this study. The simulations were done using resources of the Siberian Supercomputer Center www.sscc.ru.

We thank Eduard I. Vorobyov for constant attention to the work including careful reading and improving some chapters of the manuscript. 

\appendix

\section{Algorithm of particle initial distribution}

Here we describe the algorithm of $N$ SPH particle distribution inside the ring [$R_{min},R_{max}$]. Particles are distributed on the rings $r_i$ with equal azimuthal spacing to obtain the prescribed initial radial density in a form $\Sigma=\Sigma_0 r^{\alpha}$. 
$N=F_{ring}*N_{ring}^2$, where $F_{ring}$ is the number of particles in the first ring, and $N_{ring}$ is the number of rings. We used $F_{ring}=4$ and $N_{ring}=100$, $N_{ring}=200$, $N_{ring}=400$ to obtain a decreased, standard and increased resolution. 

$\Sigma_0=\displaystyle\frac{M_{disc} (\alpha+1)}{2 \pi (R_{max}^{\alpha+1}-R_{min}^{\alpha+1})}$. 
To satisfy the prescribed radial distribution we used recurrent formulas: 
$$r_0=R_{min}, \ \ \ r_{i+1}=\displaystyle\sqrt{r_i^2+\frac{M_{disc}(2i-1)F_{ring}}{\pi \Sigma_0 r_i^{\alpha} N}}$$ obtained from the equation
$$\displaystyle\frac{m_{SPH}F_{ring}(2i-1)}{\pi (r_i^2-r_{i-1}^2)}=\Sigma_0 r_{i-1}^{\alpha}.$$
We add the random value of uniform distribution inside the interval [$-0.001(r_i-r_{i-1})$; $0.001(r_i-r_{i-1})$] to initial radius of every particle. 

\section{Benchmarking the potential calculation method}

To verify the correctness of the potential calculation method we used the following test.

Analytical function of potential generated by a thin rectangular plate of size $2a_1\times 2a_2$ in its own plane is given by the formulae:
$$
\begin{array}{lcr}
\Phi_p (x,y)&=&-\{ I_1(a_1, x, a_2, y) + I_1(a_1, x, a_2,-y) + \\
            & &    I_2(a_1, x, a_2, y) + I_1(a_1, x, a_2,-y)\},
\end{array}
$$
where 
$$
I_1(a_1, x, a_2, y) = (a_1-x)\ln \frac{\sqrt{(a_1-x)^2 + (a_2-y)^2} +a_2-y}{\sqrt{(a_1-x)^2 + (a_2+y)^2} -a_2-y},
$$ 
and
$$
I_2(a_1, x, a_2, y) = (a_2-y)\ln \frac{\sqrt{(a_1-x)^2 + (a_2-y)^2} +a_1-x}{\sqrt{(a_1+x)^2 + (a_2-y)^2} -a_1-x}.
$$ 

Then we can calculate the potential of the same plate approximated on a grid. 
The maximum of relative difference $E(x_i, y_j) = |(\Phi_p(x_i, y_j) - \Phi_{i, j}) / \Phi_p(x_i, y_j)|$
are shown at the Table.~\ref{tab:poisson_verif}.

\begin{table}
 \caption{Maximum of the relative difference between analytical and numerical solutions for calculating the potential of a thin rectangular plate. The results are given for different grid sizes.}
 \label{tab:poisson_verif}
 \begin{tabular}{ccccc}
  \hline
  grid size & $32 \times 32$ & $128 \times 128$ &  $512 \times 512$ & $2048 \times 2048$\\
  E & $0.0627$ & $0.0178$ &  $0.0047$ & $0.0012$\\
  \hline
 \end{tabular}
\end{table}

\section{The role of $\tau_{acc}$ in accretion rate calculation}
\label{sec:TauAcc}

\begin{figure}
  \includegraphics[width=\columnwidth]{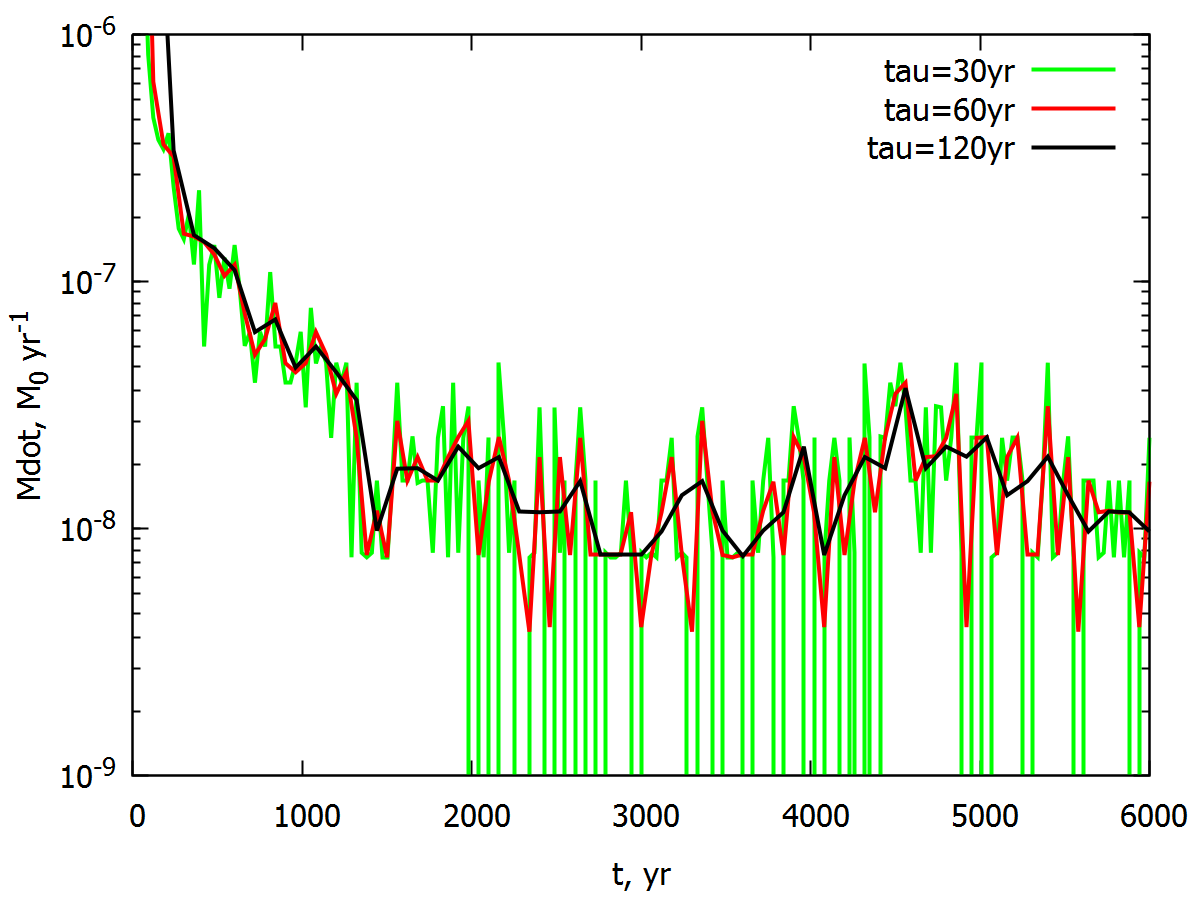} 
  \caption{The accretion rate in terms of $M_{\odot}$~yr$^{-1}$ calculated with different $\tau_{acc}$ (see Eq.(\ref{eq:Acrate})) for model 2I.}
  \label{fig:TauDependence}
   \end{figure}

\textbf{Eq.(\ref{eq:tauAc}) states that with SPH we can only study oscillations in accretion rate 
that are longer that $\tau_{min}$. However using $\tau_{acc}=\tau_{min}$ does not guarantee that 
obtained accretion rate will be free from numerical oscillations. To illustrate this statement we consider model 2I. The average accretion rate according to Figure~\ref{fig:AcRate} is about $10^{-8}M_{\odot}$~yr$^{-1}$, while $m_{SPH}=2.34 \times 10^{-7}M_{\odot}$, that provides $\tau_{min}=23.4$~yr. Figure~\ref{fig:TauDependence} demonstrates the mass accretion rate for model 2I calculated for $\tau_{acc}=30$~yr,~60~yr,~120~yr. Evidently, that green curve corresponds to value $\tau_{acc}=30$~yr has a lot of zero segments on the stage when the accretion rate is about $10^{-8}M_{\odot}$~yr$^{-1}$ and no zero segments for higher accreting period. This is in agreement with $\tau_{min}$ inversely proportional to $\dot{M}$. On the other hand, results with $\tau_{acc}=60$~yr,~120~yr has no zero segments for the whole period. Since it is easier to compare visually smoothed curves we chose $\tau_{acc}=120$~yr for general tendencies analysis presented on Figure~\ref{fig:AcRate}.} 

\section{Pattern comparison}   
\label{sec:Petterns}

\begin{figure}
  \includegraphics[width=\columnwidth]{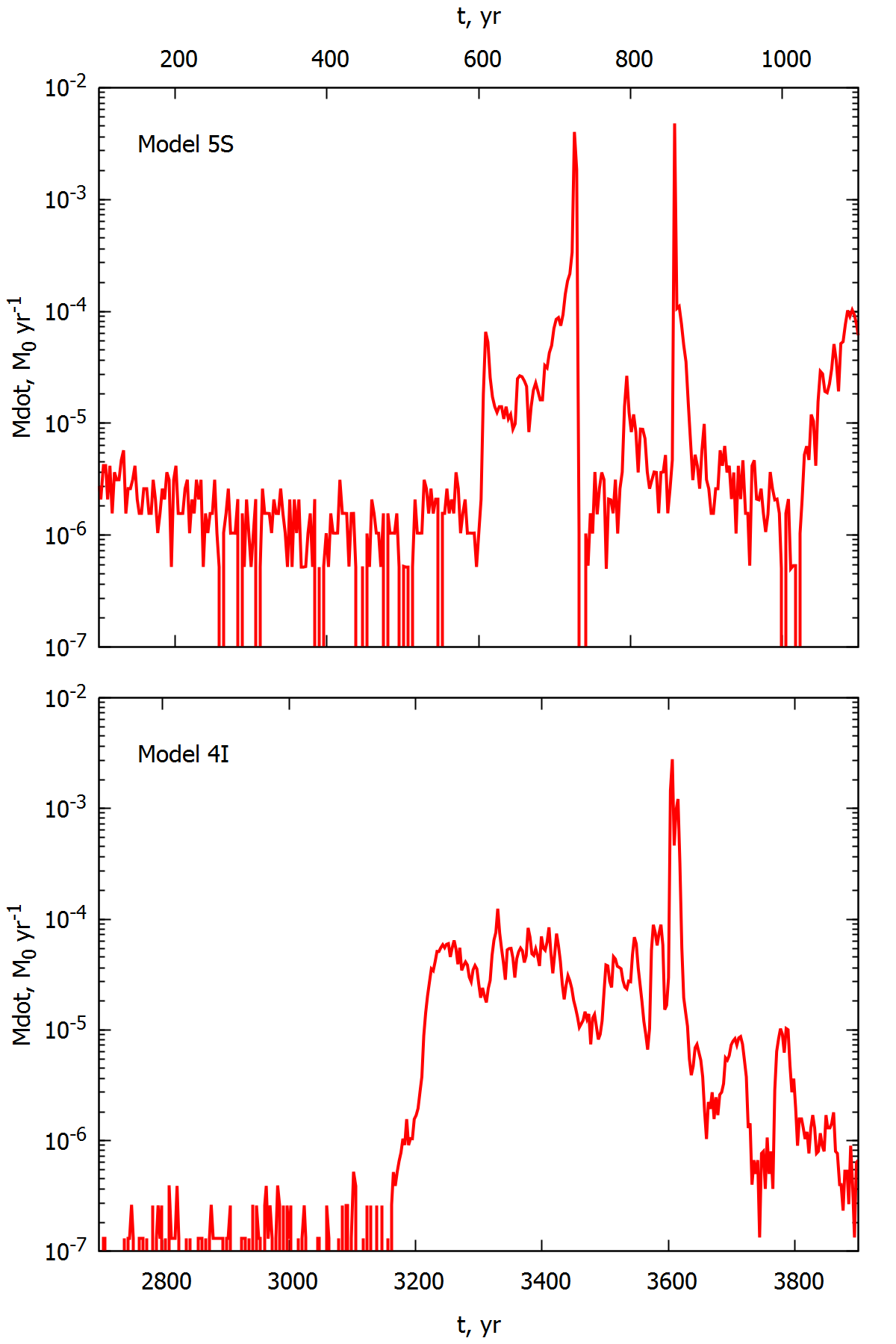} 
  \caption{The accretion rate in terms of $M_{\odot}$~yr$^{-1}$ calculated with the same $\tau=3$~yr for models 5S and 4I.}
  \label{fig:BurstPatterns}
   \end{figure}

\textbf{In this section we provide accretion history for models 4I and 5S in a form convenient for direct comparison of accretion modes. The bottom panel of Figure~{\ref{fig:BurstPatterns}} shows the accretion rate during 1200~yr for model 4I, the top panel - during 1000~yr for model 5S. Both values were found with $\tau_{acc}=3$~yr, chosen to resolve accurately periods of active accretion. The bottom panel shows clump-triggered accretion mode that features prolonged stage of elevated accretion rate, sharp burst of higher amplitude and period of increased accretion that is low than elevated value before the burst. The top panel demonstrates clearly an isolated burst after a clump-triggered one. Unlike the case of clump-triggered burst, the accretion rate before and after isolated burst is almost the same. Moreover, in both models with clump-triggered bursts the elevated value of accretion is 2-3 orders of magnitude higher than quiescent value and takes place for 100-400~yr without destruction of clump. All mentioned features allow us to differentiate these two modes of accretion.}

\section{Temperature of the clump}

\begin{figure}
  \includegraphics[width=\columnwidth]{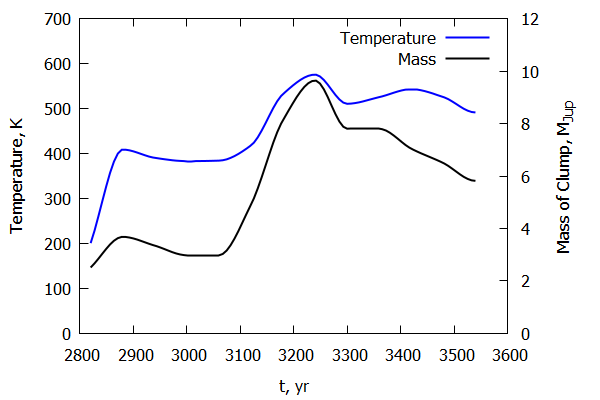} 
  \caption{Temperature (red line) and mass (blue line) of clump circled in white on Fig.  \ref{fig:ClumpBurst}.}
  \label{fig:ClumpTemp}
   \end{figure}

The temperature and the mass of the clump, responsible for generation of accretion burst in model 4I is given on Figure~\ref{fig:ClumpTemp}. The mass of the clump was calculated as a mass of all particles with density higher than 0.5 maximum density in the clump. The plotted value of temperature was taken in the center of the clump where the maximum of density (and temperature) was reached. 

\section*{References}
\bibliography{mybibfile}

\begin{thebibliography}{51}
\expandafter\ifx\csname natexlab\endcsname\relax\def\natexlab#1{#1}\fi
\providecommand{\url}[1]{\texttt{#1}}
\providecommand{\href}[2]{#2}
\providecommand{\path}[1]{#1}
\providecommand{\DOIprefix}{doi:}
\providecommand{\ArXivprefix}{arXiv:}
\providecommand{\URLprefix}{URL: }
\providecommand{\Pubmedprefix}{pmid:}
\providecommand{\doi}[1]{\href{http://dx.doi.org/#1}{\path{#1}}}
\providecommand{\Pubmed}[1]{\href{pmid:#1}{\path{#1}}}
\providecommand{\bibinfo}[2]{#2}
\ifx\xfnm\relax \def\xfnm[#1]{\unskip,\space#1}\fi
\bibitem[{{Audard} et~al.(2014){Audard}, {{\'A}brah{\'a}m}, {Dunham}, {Green},
  {Grosso}, {Hamaguchi}, {Kastner}, {K{\'o}sp{\'a}l}, {Lodato}, {Romanova},
  {Skinner}, {Vorobyov} and {Zhu}}]{Audard}
\bibinfo{author}{{Audard}, M.}, \bibinfo{author}{{{\'A}brah{\'a}m}, P.},
  \bibinfo{author}{{Dunham}, M.M.}, \bibinfo{author}{{Green}, J.D.},
  \bibinfo{author}{{Grosso}, N.}, \bibinfo{author}{{Hamaguchi}, K.},
  \bibinfo{author}{{Kastner}, J.H.}, \bibinfo{author}{{K{\'o}sp{\'a}l},
  {\'A}.}, \bibinfo{author}{{Lodato}, G.}, \bibinfo{author}{{Romanova}, M.M.},
  \bibinfo{author}{{Skinner}, S.L.}, \bibinfo{author}{{Vorobyov}, E.I.},
  \bibinfo{author}{{Zhu}, Z.}, \bibinfo{year}{2014}.
\newblock \bibinfo{title}{{Episodic Accretion in Young Stars}}.
\newblock \bibinfo{journal}{Protostars and Planets VI} ,
  \bibinfo{pages}{387--410}\DOIprefix\doi{10.2458/azu_uapress_9780816531240-ch017},
  \href{http://arxiv.org/abs/1401.3368}{\tt arXiv:1401.3368}.
\bibitem[{{Bate} and {Burkert}(1997)}]{SPHres}
\bibinfo{author}{{Bate}, M.R.}, \bibinfo{author}{{Burkert}, A.},
  \bibinfo{year}{1997}.
\newblock \bibinfo{title}{{Resolution requirements for smoothed particle
  hydrodynamics calculations with self-gravity}}.
\newblock \bibinfo{journal}{MNRAS} \bibinfo{volume}{288},
  \bibinfo{pages}{1060--1072}.
\newblock \DOIprefix\doi{10.1093/mnras/288.4.1060}.
\bibitem[{{Beck} and {Aspin}(2012)}]{BeckAspin}
\bibinfo{author}{{Beck}, T.L.}, \bibinfo{author}{{Aspin}, C.},
  \bibinfo{year}{2012}.
\newblock \bibinfo{title}{{The Nature and Evolutionary State of the FU Orionis
  Binary System}}.
\newblock \bibinfo{journal}{AJ} \bibinfo{volume}{143}, \bibinfo{pages}{55}.
\newblock \DOIprefix\doi{10.1088/0004-6256/143/3/55}.
\bibitem[{{Binney} and {Tremaine}(2008)}]{Book}
\bibinfo{author}{{Binney}, J.}, \bibinfo{author}{{Tremaine}, S.},
  \bibinfo{year}{2008}.
\newblock \bibinfo{title}{{Galactic Dynamics: Second Edition}}.
\newblock \bibinfo{publisher}{Princeton University Press}.
\bibitem[{{Boley} et~al.(2010){Boley}, {Hayfield}, {Mayer} and
  {Durisen}}]{Boley}
\bibinfo{author}{{Boley}, A.C.}, \bibinfo{author}{{Hayfield}, T.},
  \bibinfo{author}{{Mayer}, L.}, \bibinfo{author}{{Durisen}, R.H.},
  \bibinfo{year}{2010}.
\newblock \bibinfo{title}{{Clumps in the outer disk by disk instability: Why
  they are initially gas giants and the legacy of disruption}}.
\newblock \bibinfo{journal}{Icarus} \bibinfo{volume}{207},
  \bibinfo{pages}{509--516}.
\newblock \DOIprefix\doi{10.1016/j.icarus.2010.01.015},
  \href{http://arxiv.org/abs/0909.4543}{\tt arXiv:0909.4543}.
\bibitem[{{Bonnell} and {Bastien}(1992)}]{BonnellBastien}
\bibinfo{author}{{Bonnell}, I.}, \bibinfo{author}{{Bastien}, P.},
  \bibinfo{year}{1992}.
\newblock \bibinfo{title}{{A binary origin for FU Orionis stars}}.
\newblock \bibinfo{journal}{ApJL} \bibinfo{volume}{401},
  \bibinfo{pages}{L31--L34}.
\newblock \DOIprefix\doi{10.1086/186663}.
\bibitem[{{Boss}(2000)}]{Boss}
\bibinfo{author}{{Boss}, A.P.}, \bibinfo{year}{2000}.
\newblock \bibinfo{title}{{Possible Rapid Gas Giant Planet Formation in the
  Solar Nebula and Other Protoplanetary Disks}}.
\newblock \bibinfo{journal}{ApJL} \bibinfo{volume}{536},
  \bibinfo{pages}{L101--L104}.
\newblock \DOIprefix\doi{10.1086/312737}.
\bibitem[{{Dehnen} and {Aly}(2012)}]{Wendland}
\bibinfo{author}{{Dehnen}, W.}, \bibinfo{author}{{Aly}, H.},
  \bibinfo{year}{2012}.
\newblock \bibinfo{title}{{Improving convergence in smoothed particle
  hydrodynamics simulations without pairing instability}}.
\newblock \bibinfo{journal}{MNRAS} \bibinfo{volume}{425},
  \bibinfo{pages}{1068--1082}.
\newblock \DOIprefix\doi{10.1111/j.1365-2966.2012.21439.x},
  \href{http://arxiv.org/abs/1204.2471}{\tt arXiv:1204.2471}.
\bibitem[{{Durisen} et~al.(2007){Durisen}, {Boss}, {Mayer}, {Nelson}, {Quinn}
  and {Rice}}]{Durisen}
\bibinfo{author}{{Durisen}, R.H.}, \bibinfo{author}{{Boss}, A.P.},
  \bibinfo{author}{{Mayer}, L.}, \bibinfo{author}{{Nelson}, A.F.},
  \bibinfo{author}{{Quinn}, T.}, \bibinfo{author}{{Rice}, W.K.M.},
  \bibinfo{year}{2007}.
\newblock \bibinfo{title}{{Gravitational Instabilities in Gaseous
  Protoplanetary Disks and Implications for Giant Planet Formation}}.
\newblock \bibinfo{journal}{Protostars and Planets V} ,
  \bibinfo{pages}{607--622}\href{http://arxiv.org/abs/astro-ph/0603179}{\tt
  arXiv:astro-ph/0603179}.
\bibitem[{{Eastwood} and {Brownrigg}(1979)}]{eastwood}
\bibinfo{author}{{Eastwood}, J.W.}, \bibinfo{author}{{Brownrigg}, D.R.K.},
  \bibinfo{year}{1979}.
\newblock \bibinfo{title}{{Remarks on the Solution of Poisson's Equation for
  Isolated Systems}}.
\newblock \bibinfo{journal}{Journal of Computational Physics}
  \bibinfo{volume}{32}, \bibinfo{pages}{24--38}.
\newblock \DOIprefix\doi{10.1016/0021-9991(79)90139-6}.
\bibitem[{{Eisenstein} and {Hut}(1998)}]{HOP}
\bibinfo{author}{{Eisenstein}, D.J.}, \bibinfo{author}{{Hut}, P.},
  \bibinfo{year}{1998}.
\newblock \bibinfo{title}{{HOP: A New Group-Finding Algorithm for N-Body
  Simulations}}.
\newblock \bibinfo{journal}{ApJ} \bibinfo{volume}{498},
  \bibinfo{pages}{137--142}.
\newblock \DOIprefix\doi{10.1086/305535},
  \href{http://arxiv.org/abs/astro-ph/9712200}{\tt arXiv:astro-ph/9712200}.
\bibitem[{{Fridman} et~al.(1984){Fridman}, {Polyachenko}, {Aries} and
  {Poliakoff}}]{Fridman}
\bibinfo{author}{{Fridman}, A.M.}, \bibinfo{author}{{Polyachenko}, V.L.},
  \bibinfo{author}{{Aries}, A.B.}, \bibinfo{author}{{Poliakoff}, I.N.},
  \bibinfo{year}{1984}.
\newblock \bibinfo{title}{{Physics of gravitating systems. I. Equilibrium and
  stability.}}
\bibitem[{Frigo and Johnson(2005)}]{fftw}
\bibinfo{author}{Frigo, M.}, \bibinfo{author}{Johnson, S.G.},
  \bibinfo{year}{2005}.
\newblock \bibinfo{title}{The design and implementation of fftw3}.
\newblock \bibinfo{journal}{Proceedings of the IEEE} \bibinfo{volume}{93},
  \bibinfo{pages}{216--231}.
\bibitem[{{Gammie}(2001)}]{Gammie}
\bibinfo{author}{{Gammie}, C.F.}, \bibinfo{year}{2001}.
\newblock \bibinfo{title}{{Nonlinear Outcome of Gravitational Instability in
  Cooling, Gaseous Disks}}.
\newblock \bibinfo{journal}{ApJ} \bibinfo{volume}{553},
  \bibinfo{pages}{174--183}.
\newblock \DOIprefix\doi{10.1086/320631},
  \href{http://arxiv.org/abs/astro-ph/0101501}{\tt arXiv:astro-ph/0101501}.
\bibitem[{{Gingold} and {Monaghan}(1977)}]{GMSPH}
\bibinfo{author}{{Gingold}, R.A.}, \bibinfo{author}{{Monaghan}, J.J.},
  \bibinfo{year}{1977}.
\newblock \bibinfo{title}{{Smoothed particle hydrodynamics - Theory and
  application to non-spherical stars}}.
\newblock \bibinfo{journal}{MNRAS} \bibinfo{volume}{181},
  \bibinfo{pages}{375--389}.
\newblock \DOIprefix\doi{10.1093/mnras/181.3.375}.
\bibitem[{{Hockney} and {Eastwood}(1981)}]{Hockney}
\bibinfo{author}{{Hockney}, R.W.}, \bibinfo{author}{{Eastwood}, J.W.},
  \bibinfo{year}{1981}.
\newblock \bibinfo{title}{{Computer Simulation Using Particles}}.
\bibitem[{{Lodato} and {Clarke}(2011)}]{LodatoClarke}
\bibinfo{author}{{Lodato}, G.}, \bibinfo{author}{{Clarke}, C.J.},
  \bibinfo{year}{2011}.
\newblock \bibinfo{title}{{Resolution requirements for smoothed particle
  hydrodynamics simulations of self-gravitating accretion discs}}.
\newblock \bibinfo{journal}{MNRAS} \bibinfo{volume}{413},
  \bibinfo{pages}{2735--2740}.
\newblock \DOIprefix\doi{10.1111/j.1365-2966.2011.18344.x},
  \href{http://arxiv.org/abs/1101.2448}{\tt arXiv:1101.2448}.
\bibitem[{{Lucy}(1977)}]{Lucy}
\bibinfo{author}{{Lucy}, L.B.}, \bibinfo{year}{1977}.
\newblock \bibinfo{title}{{A numerical approach to the testing of the fission
  hypothesis}}.
\newblock \bibinfo{journal}{AJ} \bibinfo{volume}{82},
  \bibinfo{pages}{1013--1024}.
\newblock \DOIprefix\doi{10.1086/112164}.
\bibitem[{{Machida} et~al.(2011){Machida}, {Inutsuka} and
  {Matsumoto}}]{MachidaAc}
\bibinfo{author}{{Machida}, M.N.}, \bibinfo{author}{{Inutsuka}, S.i.},
  \bibinfo{author}{{Matsumoto}, T.}, \bibinfo{year}{2011}.
\newblock \bibinfo{title}{{Recurrent Planet Formation and Intermittent
  Protostellar Outflows Induced by Episodic Mass Accretion}}.
\newblock \bibinfo{journal}{ApJ} \bibinfo{volume}{729}, \bibinfo{pages}{42}.
\newblock \DOIprefix\doi{10.1088/0004-637X/729/1/42},
  \href{http://arxiv.org/abs/1101.1997}{\tt arXiv:1101.1997}.
\bibitem[{{Meru} and {Bate}(2012)}]{MeruBate}
\bibinfo{author}{{Meru}, F.}, \bibinfo{author}{{Bate}, M.R.},
  \bibinfo{year}{2012}.
\newblock \bibinfo{title}{{On the convergence of the critical cooling
  time-scale for the fragmentation of self-gravitating discs}}.
\newblock \bibinfo{journal}{MNRAS} \bibinfo{volume}{427},
  \bibinfo{pages}{2022--2046}.
\newblock \DOIprefix\doi{10.1111/j.1365-2966.2012.22035.x},
  \href{http://arxiv.org/abs/1209.1107}{\tt arXiv:1209.1107}.
\bibitem[{{Monaghan}(1992)}]{SPH}
\bibinfo{author}{{Monaghan}, J.J.}, \bibinfo{year}{1992}.
\newblock \bibinfo{title}{{Smoothed particle hydrodynamics}}.
\newblock \bibinfo{journal}{ARA\&A} \bibinfo{volume}{30},
  \bibinfo{pages}{543--574}.
\newblock \DOIprefix\doi{10.1146/annurev.aa.30.090192.002551}.
\bibitem[{{Nayakshin}(2010)}]{Nayakshin}
\bibinfo{author}{{Nayakshin}, S.}, \bibinfo{year}{2010}.
\newblock \bibinfo{title}{{Grain sedimentation inside giant planet embryos}}.
\newblock \bibinfo{journal}{MNRAS} \bibinfo{volume}{408},
  \bibinfo{pages}{2381--2396}.
\newblock \DOIprefix\doi{10.1111/j.1365-2966.2010.17289.x},
  \href{http://arxiv.org/abs/1007.4162}{\tt arXiv:1007.4162}.
\bibitem[{{Nayakshin} and {Cha}(2013)}]{NayakshinCha}
\bibinfo{author}{{Nayakshin}, S.}, \bibinfo{author}{{Cha}, S.H.},
  \bibinfo{year}{2013}.
\newblock \bibinfo{title}{{Radiative feedback from protoplanets in
  self-gravitating protoplanetary discs}}.
\newblock \bibinfo{journal}{MNRAS} \bibinfo{volume}{435},
  \bibinfo{pages}{2099--2108}.
\newblock \DOIprefix\doi{10.1093/mnras/stt1426},
  \href{http://arxiv.org/abs/1306.4924}{\tt arXiv:1306.4924}.
\bibitem[{{Nayakshin} and {Lodato}(2012)}]{NayakshinLodato}
\bibinfo{author}{{Nayakshin}, S.}, \bibinfo{author}{{Lodato}, G.},
  \bibinfo{year}{2012}.
\newblock \bibinfo{title}{{Fu Ori outbursts and the planet-disc mass
  exchange}}.
\newblock \bibinfo{journal}{MNRAS} \bibinfo{volume}{426},
  \bibinfo{pages}{70--90}.
\newblock \DOIprefix\doi{10.1111/j.1365-2966.2012.21612.x},
  \href{http://arxiv.org/abs/1110.6316}{\tt arXiv:1110.6316}.
\bibitem[{{Nelson}(2006)}]{Nelson}
\bibinfo{author}{{Nelson}, A.F.}, \bibinfo{year}{2006}.
\newblock \bibinfo{title}{{Numerical requirements for simulations of
  self-gravitating and non-self-gravitating discs}}.
\newblock \bibinfo{journal}{MNRAS} \bibinfo{volume}{373},
  \bibinfo{pages}{1039--1073}.
\newblock \DOIprefix\doi{10.1111/j.1365-2966.2006.11119.x},
  \href{http://arxiv.org/abs/astro-ph/0609493}{\tt arXiv:astro-ph/0609493}.
\bibitem[{{Paardekooper}(2012)}]{Paardekooper}
\bibinfo{author}{{Paardekooper}, S.J.}, \bibinfo{year}{2012}.
\newblock \bibinfo{title}{{Numerical convergence in self-gravitating shearing
  sheet simulations and the stochastic nature of disc fragmentation}}.
\newblock \bibinfo{journal}{MNRAS} \bibinfo{volume}{421},
  \bibinfo{pages}{3286--3299}.
\newblock \DOIprefix\doi{10.1111/j.1365-2966.2012.20553.x},
  \href{http://arxiv.org/abs/1201.3371}{\tt arXiv:1201.3371}.
\bibitem[{{Pfalzner}(2008)}]{Pfalzner2008}
\bibinfo{author}{{Pfalzner}, S.}, \bibinfo{year}{2008}.
\newblock \bibinfo{title}{{Encounter-driven accretion in young stellar clusters
  - A connection to FUors?}}
\newblock \bibinfo{journal}{A\&A} \bibinfo{volume}{492},
  \bibinfo{pages}{735--741}.
\newblock \DOIprefix\doi{10.1051/0004-6361:200810879},
  \href{http://arxiv.org/abs/0810.2854}{\tt arXiv:0810.2854}.
\bibitem[{{Pickett} et~al.(1998){Pickett}, {Cassen}, {Durisen} and
  {Link}}]{Pickett1998}
\bibinfo{author}{{Pickett}, B.K.}, \bibinfo{author}{{Cassen}, P.},
  \bibinfo{author}{{Durisen}, R.H.}, \bibinfo{author}{{Link}, R.},
  \bibinfo{year}{1998}.
\newblock \bibinfo{title}{{The Effects of Thermal Energetics on
  Three-dimensional Hydrodynamic Instabilities in Massive Protostellar Disks}}.
\newblock \bibinfo{journal}{ApJ} \bibinfo{volume}{504},
  \bibinfo{pages}{468--491}.
\newblock \DOIprefix\doi{10.1086/306059}.
\bibitem[{{Pickett} et~al.(2000){Pickett}, {Cassen}, {Durisen} and
  {Link}}]{Pickett2000}
\bibinfo{author}{{Pickett}, B.K.}, \bibinfo{author}{{Cassen}, P.},
  \bibinfo{author}{{Durisen}, R.H.}, \bibinfo{author}{{Link}, R.},
  \bibinfo{year}{2000}.
\newblock \bibinfo{title}{{The Effects of Thermal Energetics on
  Three-dimensional Hydrodynamic Instabilities in Massive Protostellar Disks.
  II. High-Resolution and Adiabatic Evolutions}}.
\newblock \bibinfo{journal}{ApJ} \bibinfo{volume}{529},
  \bibinfo{pages}{1034--1053}.
\newblock \DOIprefix\doi{10.1086/308301}.
\bibitem[{{Price}(2012)}]{Price}
\bibinfo{author}{{Price}, D.J.}, \bibinfo{year}{2012}.
\newblock \bibinfo{title}{{Smoothed particle hydrodynamics and
  magnetohydrodynamics}}.
\newblock \bibinfo{journal}{Journal of Computational Physics}
  \bibinfo{volume}{231}, \bibinfo{pages}{759--794}.
\newblock \DOIprefix\doi{10.1016/j.jcp.2010.12.011},
  \href{http://arxiv.org/abs/1012.1885}{\tt arXiv:1012.1885}.
\bibitem[{{Reipurth} and {Aspin}(2004)}]{ReipurthAspin}
\bibinfo{author}{{Reipurth}, B.}, \bibinfo{author}{{Aspin}, C.},
  \bibinfo{year}{2004}.
\newblock \bibinfo{title}{{The FU Orionis Binary System and the Formation of
  Close Binaries}}.
\newblock \bibinfo{journal}{ApJL} \bibinfo{volume}{608},
  \bibinfo{pages}{L65--L68}.
\newblock \DOIprefix\doi{10.1086/422250}.
\bibitem[{{Rice} et~al.(2012){Rice}, {Forgan} and {Armitage}}]{RiceCool}
\bibinfo{author}{{Rice}, W.K.M.}, \bibinfo{author}{{Forgan}, D.H.},
  \bibinfo{author}{{Armitage}, P.J.}, \bibinfo{year}{2012}.
\newblock \bibinfo{title}{{Convergence of smoothed particle hydrodynamics
  simulations of self-gravitating accretion discs: sensitivity to the
  implementation of radiative cooling}}.
\newblock \bibinfo{journal}{MNRAS} \bibinfo{volume}{420},
  \bibinfo{pages}{1640--1647}.
\newblock \DOIprefix\doi{10.1111/j.1365-2966.2011.20153.x},
  \href{http://arxiv.org/abs/1111.3147}{\tt arXiv:1111.3147}.
\bibitem[{{Schuessler} and {Schmitt}(1981)}]{Pairing}
\bibinfo{author}{{Schuessler}, I.}, \bibinfo{author}{{Schmitt}, D.},
  \bibinfo{year}{1981}.
\newblock \bibinfo{title}{{Comments on smoothed particle hydrodynamics}}.
\newblock \bibinfo{journal}{A\&A} \bibinfo{volume}{97},
  \bibinfo{pages}{373--379}.
\bibitem[{{Snytnikov} and
  {Stoyanovskaya}(2016)}]{SnytnikovStoyanovskaya2016CMP}
\bibinfo{author}{{Snytnikov}, V.N.}, \bibinfo{author}{{Stoyanovskaya}, O.P.},
  \bibinfo{year}{2016}.
\newblock \bibinfo{title}{{On the correctness of numerical simulation of
  gravitational instability with the evolution of multiple gravitational
  collapses}}.
\newblock \bibinfo{journal}{Computational Methods and Programming}
  \bibinfo{volume}{17}, \bibinfo{pages}{365--379, in Russian.}
\bibitem[{{Springel}(2005)}]{Gadget2}
\bibinfo{author}{{Springel}, V.}, \bibinfo{year}{2005}.
\newblock \bibinfo{title}{{The cosmological simulation code GADGET-2}}.
\newblock \bibinfo{journal}{MNRAS} \bibinfo{volume}{364},
  \bibinfo{pages}{1105--1134}.
\newblock \DOIprefix\doi{10.1111/j.1365-2966.2005.09655.x},
  \href{http://arxiv.org/abs/astro-ph/0505010}{\tt arXiv:astro-ph/0505010}.
\bibitem[{{Springel}(2010)}]{SpringelSPH}
\bibinfo{author}{{Springel}, V.}, \bibinfo{year}{2010}.
\newblock \bibinfo{title}{{Smoothed Particle Hydrodynamics in Astrophysics}}.
\newblock \bibinfo{journal}{ARA\&A} \bibinfo{volume}{48},
  \bibinfo{pages}{391--430}.
\newblock \DOIprefix\doi{10.1146/annurev-astro-081309-130914},
  \href{http://arxiv.org/abs/1109.2219}{\tt arXiv:1109.2219}.
\bibitem[{{Stoyanovskaya}(2016)}]{Stoyanovskaya2016CMP}
\bibinfo{author}{{Stoyanovskaya}, O.P.}, \bibinfo{year}{2016}.
\newblock \bibinfo{title}{{Numerical simulation of gravitational instability
  development and clump formation in massive circumstellar disks using integral
  characteristics for the interpretation of results}}.
\newblock \bibinfo{journal}{Computational Methods and Programming}
  \bibinfo{volume}{17}, \bibinfo{pages}{339--352, in Russian.}
\bibitem[{{Takahashi} et~al.(2016){Takahashi}, {Tsukamoto} and
  {Inutsuka}}]{Takahashi}
\bibinfo{author}{{Takahashi}, S.Z.}, \bibinfo{author}{{Tsukamoto}, Y.},
  \bibinfo{author}{{Inutsuka}, S.}, \bibinfo{year}{2016}.
\newblock \bibinfo{title}{{A revised condition for self-gravitational
  fragmentation of protoplanetary discs}}.
\newblock \bibinfo{journal}{MNRAS} \bibinfo{volume}{458},
  \bibinfo{pages}{3597--3612}.
\newblock \DOIprefix\doi{10.1093/mnras/stw557},
  \href{http://arxiv.org/abs/1603.01402}{\tt arXiv:1603.01402}.
\bibitem[{{Thacker} and {Couchman}(2006)}]{Hydra}
\bibinfo{author}{{Thacker}, R.J.}, \bibinfo{author}{{Couchman}, H.M.P.},
  \bibinfo{year}{2006}.
\newblock \bibinfo{title}{{A parallel adaptive P$^{3}$M code with hierarchical
  particle reordering}}.
\newblock \bibinfo{journal}{Computer Physics Communications}
  \bibinfo{volume}{174}, \bibinfo{pages}{540--554}.
\newblock \DOIprefix\doi{10.1016/j.cpc.2005.12.001},
  \href{http://arxiv.org/abs/astro-ph/0512030}{\tt arXiv:astro-ph/0512030}.
\bibitem[{{Thies} et~al.(2010){Thies}, {Kroupa}, {Goodwin}, {Stamatellos} and
  {Whitworth}}]{Thies2010}
\bibinfo{author}{{Thies}, I.}, \bibinfo{author}{{Kroupa}, P.},
  \bibinfo{author}{{Goodwin}, S.P.}, \bibinfo{author}{{Stamatellos}, D.},
  \bibinfo{author}{{Whitworth}, A.P.}, \bibinfo{year}{2010}.
\newblock \bibinfo{title}{{Tidally Induced Brown Dwarf and Planet Formation in
  Circumstellar Disks}}.
\newblock \bibinfo{journal}{ApJ} \bibinfo{volume}{717},
  \bibinfo{pages}{577--585}.
\newblock \DOIprefix\doi{10.1088/0004-637X/717/1/577},
  \href{http://arxiv.org/abs/1005.3017}{\tt arXiv:1005.3017}.
\bibitem[{{Thomas} and {Couchman}(1992)}]{TC1992}
\bibinfo{author}{{Thomas}, P.A.}, \bibinfo{author}{{Couchman}, H.M.P.},
  \bibinfo{year}{1992}.
\newblock \bibinfo{title}{{Simulating the formation of a cluster of galaxies}}.
\newblock \bibinfo{journal}{MNRAS} \bibinfo{volume}{257},
  \bibinfo{pages}{11--31}.
\newblock \DOIprefix\doi{10.1093/mnras/257.1.11}.
\bibitem[{{Truelove} et~al.(1997){Truelove}, {Klein}, {McKee}, {Holliman},
  {Howell} and {Greenough}}]{Truelove}
\bibinfo{author}{{Truelove}, J.K.}, \bibinfo{author}{{Klein}, R.I.},
  \bibinfo{author}{{McKee}, C.F.}, \bibinfo{author}{{Holliman}, II, J.H.},
  \bibinfo{author}{{Howell}, L.H.}, \bibinfo{author}{{Greenough}, J.A.},
  \bibinfo{year}{1997}.
\newblock \bibinfo{title}{{The Jeans Condition: A New Constraint on Spatial
  Resolution in Simulations of Isothermal Self-gravitational Hydrodynamics}}.
\newblock \bibinfo{journal}{ApJL} \bibinfo{volume}{489},
  \bibinfo{pages}{L179--L183}.
\newblock \DOIprefix\doi{10.1086/310975}.
\bibitem[{{Vorobyov}(2010)}]{Vorobyov2010}
\bibinfo{author}{{Vorobyov}, E.I.}, \bibinfo{year}{2010}.
\newblock \bibinfo{title}{{Embedded Protostellar Disks Around (Sub-)Solar
  Protostars. I. Disk Structure and Evolution}}.
\newblock \bibinfo{journal}{ApJ} \bibinfo{volume}{723},
  \bibinfo{pages}{1294--1307}.
\newblock \DOIprefix\doi{10.1088/0004-637X/723/2/1294},
  \href{http://arxiv.org/abs/1009.2073}{\tt arXiv:1009.2073}.
\bibitem[{{Vorobyov}(2013)}]{Vorobyov2013}
\bibinfo{author}{{Vorobyov}, E.I.}, \bibinfo{year}{2013}.
\newblock \bibinfo{title}{{Formation of giant planets and brown dwarfs on wide
  orbits}}.
\newblock \bibinfo{journal}{A\&A} \bibinfo{volume}{552}, \bibinfo{pages}{A129}.
\newblock \DOIprefix\doi{10.1051/0004-6361/201220601},
  \href{http://arxiv.org/abs/1302.1892}{\tt arXiv:1302.1892}.
\bibitem[{{Vorobyov} and {Basu}(2006)}]{VorobyovBasu2006}
\bibinfo{author}{{Vorobyov}, E.I.}, \bibinfo{author}{{Basu}, S.},
  \bibinfo{year}{2006}.
\newblock \bibinfo{title}{{The Burst Mode of Protostellar Accretion}}.
\newblock \bibinfo{journal}{ApJ} \bibinfo{volume}{650},
  \bibinfo{pages}{956--969}.
\newblock \DOIprefix\doi{10.1086/507320},
  \href{http://arxiv.org/abs/astro-ph/0607118}{\tt arXiv:astro-ph/0607118}.
\bibitem[{{Vorobyov} and {Basu}(2008)}]{VorobyovBasu2008}
\bibinfo{author}{{Vorobyov}, E.I.}, \bibinfo{author}{{Basu}, S.},
  \bibinfo{year}{2008}.
\newblock \bibinfo{title}{{Mass Accretion Rates in Self-Regulated Disks of T
  Tauri Stars}}.
\newblock \bibinfo{journal}{ApJL} \bibinfo{volume}{676}, \bibinfo{pages}{L139}.
\newblock \DOIprefix\doi{10.1086/587514},
  \href{http://arxiv.org/abs/0802.2242}{\tt arXiv:0802.2242}.
\bibitem[{{Vorobyov} and {Basu}(2010)}]{VorobyovBasu2010}
\bibinfo{author}{{Vorobyov}, E.I.}, \bibinfo{author}{{Basu}, S.},
  \bibinfo{year}{2010}.
\newblock \bibinfo{title}{{The Burst Mode of Accretion and Disk Fragmentation
  in the Early Embedded Stages of Star Formation}}.
\newblock \bibinfo{journal}{ApJ} \bibinfo{volume}{719},
  \bibinfo{pages}{1896--1911}.
\newblock \DOIprefix\doi{10.1088/0004-637X/719/2/1896},
  \href{http://arxiv.org/abs/1007.2993}{\tt arXiv:1007.2993}.
\bibitem[{{Vorobyov} and {Basu}(2015)}]{VorobyovBasu2015}
\bibinfo{author}{{Vorobyov}, E.I.}, \bibinfo{author}{{Basu}, S.},
  \bibinfo{year}{2015}.
\newblock \bibinfo{title}{{Variable Protostellar Accretion with Episodic
  Bursts}}.
\newblock \bibinfo{journal}{ApJ} \bibinfo{volume}{805}, \bibinfo{pages}{17}.
\newblock \DOIprefix\doi{10.1088/0004-637X/805/2/115},
  \href{http://arxiv.org/abs/1503.07888}{\tt arXiv:1503.07888}.
\bibitem[{{Wang} et~al.(2004){Wang}, {Apai}, {Henning} and {Pascucci}}]{Wang}
\bibinfo{author}{{Wang}, H.}, \bibinfo{author}{{Apai}, D.},
  \bibinfo{author}{{Henning}, T.}, \bibinfo{author}{{Pascucci}, I.},
  \bibinfo{year}{2004}.
\newblock \bibinfo{title}{{FU Orionis: A Binary Star?}}
\newblock \bibinfo{journal}{ApJL} \bibinfo{volume}{601},
  \bibinfo{pages}{L83--L86}.
\newblock \DOIprefix\doi{10.1086/381705},
  \href{http://arxiv.org/abs/astro-ph/0311606}{\tt arXiv:astro-ph/0311606}.
\bibitem[{{Young} and {Clarke}(2015)}]{YoungClarke}
\bibinfo{author}{{Young}, M.D.}, \bibinfo{author}{{Clarke}, C.J.},
  \bibinfo{year}{2015}.
\newblock \bibinfo{title}{{Dependence of fragmentation in self-gravitating
  accretion discs on small-scale structure}}.
\newblock \bibinfo{journal}{MNRAS} \bibinfo{volume}{451},
  \bibinfo{pages}{3987--3994}.
\newblock \DOIprefix\doi{10.1093/mnras/stv1266},
  \href{http://arxiv.org/abs/1506.02560}{\tt arXiv:1506.02560}.
\bibitem[{{Zhu} et~al.(2012){Zhu}, {Hartmann}, {Nelson} and
  {Gammie}}]{ZhuClumps}
\bibinfo{author}{{Zhu}, Z.}, \bibinfo{author}{{Hartmann}, L.},
  \bibinfo{author}{{Nelson}, R.P.}, \bibinfo{author}{{Gammie}, C.F.},
  \bibinfo{year}{2012}.
\newblock \bibinfo{title}{{Challenges in Forming Planets by Gravitational
  Instability: Disk Irradiation and Clump Migration, Accretion, and Tidal
  Destruction}}.
\newblock \bibinfo{journal}{ApJ} \bibinfo{volume}{746}, \bibinfo{pages}{110}.
\newblock \DOIprefix\doi{10.1088/0004-637X/746/1/110},
  \href{http://arxiv.org/abs/1111.6943}{\tt arXiv:1111.6943}.

\end{thebibliography}

\end{document}